\documentclass{aa}  

\usepackage{amsmath, amssymb, graphicx,txfonts,hyperref}
\usepackage{xcolor}
\begin{document}

\title{Searching for new hypervelocity stars with $Gaia$ DR3 and VLT/FORS2 spectroscopy\thanks{Based on observations collected at the European Southern Observatory under  ESO programs 110.23UC.007 and 111.24P0.006.}}

\author{Julio A. Carballo-Bello\inst{1\text{*}}, Pau Ramos \inst{2}, Jes\'us M. Corral-Santana \inst{3}, Camila Navarrete \inst{4}, Teresa Antoja \inst{5, 6, 7}, \\
Felipe Gran \inst{4}, Mat\'ias Bla\~na \inst{8} and Marcelo D. Mora\inst{9}}

\institute{Instituto de Alta Investigaci\'{o}n, Universidad de Tarapac\'{a}, Casilla 7D, Arica, Chile
\and
National Astronomical Observatory of Japan, Mitaka-shi, Tokyo 181-8588, Japan
\and
European Southern Observatory, Alonso de C\'ordova 3107, Vitacura, Santiago, Chile
\and
Universit{\'e} C{\^o}te d'Azur, Observatoire de la C{\^o}te d'Azur, CNRS, Laboratoire Lagrange, Bd de l'Observatoire, CS 34229, 06304, Nice Cedex 4, France
\and
Institut de Ci\`encies del Cosmos (ICCUB), Universitat de Barcelona (UB), Mart\'{i} i Franqu\`{e}s 1, E-08028 Barcelona, Spain
\and
Departament de Física Qu\`antica i Astrofísica (FQA), Universitat de Barcelona (UB), Mart\'{i} i Franqu\`{e}s 1, E-08028 Barcelona, Spain
\and
Institut d'Estudis Espacials de Catalunya (IEEC), c. Gran Capit\`a, 2-4, E-08034 Barcelona, Spain
\and
Vicerrector\'ia de Investigaci\'on y Postgrado, Universidad de La Serena, La Serena 1700000, Chile
\and
Las Campanas Observatory, Carnegie Observatories, Casilla 601, La Serena, 7820436, Chile\\
\text{*}\email{jcarballo@academicos.uta.cl}\\
}

\date{Received March 27, 2025; accepted June 26, 2025}

  \abstract
   {Hypervelocity stars are unique objects moving through the Milky Way at speeds exceeding the local escape velocity, providing valuable insights into the Galaxy’s gravitational potential and the properties of its central supermassive black hole. The advent of {\it Gaia} DR3 offers an unprecedented astrometric precision, enabling the discovery of new hypervelocity stars and facilitating their characterization.}
   {This study seeks to identify and characterize hypervelocity star candidates using {\it Gaia} DR3 data, focusing on stars lacking radial velocity measurements. Our goal was to estimate the total velocities of these stars and establish their origin within the Galactic framework, if possible.}
   {We applied strict selection criteria to {\it Gaia} DR3 data, focusing on sources with low parallax uncertainties and high astrometric fidelity. The distributions of the total velocities in the Galactic rest frame were derived and used to identify candidates. Spectroscopic follow-up with VLT/FORS2 provided radial velocity measurements for a subset of these candidates. We evaluated the probabilities of stars exceeding local escape velocities under different Galactic potential models and traced their past orbits to identify possible origins.}
   {From {\it Gaia} DR3, we identified 149 hypervelocity star candidates with probabilities $P_{\rm esc} \geq 50\%$ of exceeding local escape velocities. Our follow-up spectroscopy for 23 of those sources confirms that the selected targets are traveling at high velocities, with many appearing to escape the Galaxy, depending on the Galactic potential adopted. We found that, except for one target with a minimum distance of $\sim$1\,kpc within uncertainties, none of them seem to have originated at the Galactic center. On the other hand, our analysis suggests that nearly one-third of the stars may have an extra-Galactic origin. These findings highlight the need for more precise astrometric and spectroscopic data to conclusively determine the origins of hypervelocity stars and improve models of the Galactic potential.}
   {}
   \keywords{Galaxy:general --
                Stars: kinematics and dynamics -- astrometry
               }
\authorrunning{Carballo-Bello et al.}
\titlerunning{Hypervelocity stars with {\it Gaia} DR3 and FORS2 spectroscopy}

   \maketitle

\section{Introduction}

Hypervelocity stars (HVSs), which are traveling across our Galaxy at velocities above the local escape velocity ($v_{\rm esc}$), have attracted significant attention, as evidenced by the volume of recent studies focused on these objects. This interest has been driven by the arrival of wide-sky surveys, which provide the necessary data to characterize the orbits of millions of stars with unprecedented precision \citep{Brown2015}. Between the pioneering work of \cite{Hills1988} and the identification of the first HVS - SDSS J090745.0+02450 \citep{Brown2005} - the field saw mostly theoretical developments. However, in the last decade, the known Galactic  population of high-velocity (HiVel) stars has increased by hundreds, including several dozen confirmed HVSs. Although there is no universally adopted definition of a HiVel star, they are generally considered to be objects with total Galactocentric velocities above $\sim 300 - 400$\,km\,s$^{-1}$ \citep[e.g.][]{Du2018,Hattori2018,Quispe2022}.

The main mechanism for generating HVSs is the so-called Hills mechanism \citep{Hills1988,Yu2003}, which predicts that one of the members in a binary star system might be ejected and reach such velocities during the interaction with the supermassive black hole (SMBH) at the center of our Galaxy \citep{EHT}. However, only a few confirmed HVSs appear to originate from the inner regions of the Milky Way (MW) when their orbits are traced back in time \citep[e.g.][]{Koposov2020}. Interestingly, \cite{Chu2023} found a lower binary fraction in the surroundings of the Galactic SMBH (47\% versus 70\% in the field), which is consistent with the scenario in which the SMBH is playing a relevant role in the disruption of binary systems. Although some bound HiVels originating from the MV center have been reported in the literature \citep[e.g.][]{Hattori2025}, we reserve the term HVS in this work exclusively to refer to unbound stars. \cite{Liao2023} proposed that a close encounter between a single star and the SMBH might also produce a HVS, but in a regime of rates and/or luminosities that makes their detection more difficult. Orbitally decayed globular clusters may also contribute with HVSs after a close encounter with the central SMBH \citep{Capuzzo-Dolcetta2015,Fragione2016}. Numerical simulations also show that intermediate-mass black holes (IMBHs) sinking to the center of the Galaxy may accelerate stars \citep{Baumgardt2006}.

Alternatively, HVSs might  originate in the disruption of  accreted MW satellite dwarf galaxies \citep{Abadi2009,Piffl2011}, where stars are stripped from their progenitor galaxy during its pericentric passage, close to the Galactic center, and pushed into high-velocity orbits. Indeed, recent studies \citep{Li2022,Huang2021} have shown that $\sim$ 60 HiVels, some of them possibly HVSs, were originated in the Sagittarius dwarf galaxy \citep[see also][]{Du2018,Du2019,Montanari2019}. Even the Magellanic Clouds seem to be a tentative birth place for those runners, since a few distant stars have past orbits pointing towards them \citep{Edelmann2005,Irrgang2018,Erkal2019,Lin2023}. Recently, it has been proved that it is theoretically possible for a HVS originated in the Andromeda galaxy to reach the MW \citep{Gulzow2023}. Other ``violent'' scenarios, such as a supernova explosion in a binary system where the companion star is ejected, may also produce HVSs although most of them are HiVels, with total velocities of a few hundred km\,s$^{-1}$ \citep[e.g.][]{Irrgang2021,Ruiz-Lapuente2023}. The latter process is possibly responsible for the so-called runaway stars, which are ejected from the Galactic disk without interaction with the SMBH \citep{Silva2011}.

The search for such peculiar stars is of interest not only because of the discovery itself, but to understand the nature of the central SMBH and the mass distribution of the sections of the MW they cross \citep[e.g. see ][and references therein]{Gallo2022}. Since the expected number \citep[][]{Yu2003,Brown2014} and the fraction of HVSs generated via the different mechanisms described above is unknown, the search has continued since the one discovered by \cite{Brown2005} through dedicated (or not) observational campaigns, which have revealed the presence of tens of HVSs in the Galactic halo \citep[e.g.][]{Hirsch2005,Brown2006,Gualandris2007,Tillich2009,Brown2010,Brown2014,Koposov2020,Burgasser2024,Verberne2024}. 

The arrival of the European Space Mission {\it Gaia} \citep{Gaia2016,Gaia2018,Gaia2023} represents a revolution for this field, since for the first time we have access to precise astrometric information (coordinates, distances, proper motions) and radial velocities ($v_{\rm los}$) for a significant number of sources in our Galaxy. This unprecedented dataset enables a detailed study of the past and present orbits of HVS candidates. In this context, the identification of more of these travelers - excellent stellar probes from the inner regions of the MW \citep{Contigiani2019,Evans2022} - was among the many potential applications proposed for {\it Gaia} datasets following its first data release \cite[see][]{Marchetti2017,Marchetti2018}. Since then, the successive Gaia data releases have been mined in the search for HVS candidates \citep[e.g.][]{Du2018,Generozov2020,Li2022,Liao2023,Marchetti2022,Li2021,Marchetti2019,Boubert2018,Scholz2024}.

Of course, despite the tremendous advancement in the discovery and study of Galactic HVSs, the uncertainties associated to the astrometric parameters are still affecting our ability to identify these fast travelers. Indeed, inferring the location within the Galaxy where these stars gained enough energy to escape it is still hard for a significant fraction of HVS candidates. In the case of Gaia, the parallax and proper motions errors for faint sources, together with the lack of measured $v_{\rm los}$ for most of the objects, is hampering our attempts to find new HVSs, and more importantly, compute their past orbits to study the importance of the different ejection mechanisms proposed.   

With this paper, we intend to contribute to the census of known Galactic unbound HVS candidates, by combining Gaia data with ground-based follow-up spectroscopy for those sources which are likely moving at velocities above $v_{\rm esc}$. This will allow us not only to confirm their total velocities but to search for their ejection location within the MW. 

\section{Methodology. Sample of HVS candidates}
\label{methodology}
Our main goal with this work was to identify HVS candidates whose orbits  cannot be reconstructed due to the lack of $v_{\rm los}$. Therefore, we exclude from our analysis all sources with $v_{\rm los}$ already provided by {\it Gaia} and will assume that the hypothetical HVSs in {\it Gaia} DR3 have already been identified in previous studies if their $v_{\rm los}$ were known. Here, we focus on high-quality astrometry sources with incomplete Gaia information in the 6D phase space.

First, we restrict our sample to those stars whose parallaxes have relatively low associated uncertainties by only including sources with $\sigma_{\varpi} / \varpi \leq 0.2$. This limit, combined with additional quality cuts, allows us to derive heliocentric distances from the inversion of parallaxes alone, although we have checked that other estimates \citep[e.g.][]{Bailer-Jones2021} are consistent within errors. We also used two parameters that account for the quality of the parameters provided by Gaia for a given source: the Renormalized Unit Weight Error \citep[RUWE; see description in ][]{Lindegren2021} for which we set the usual limit of RUWE $<$ 1.4, and the astrometric fidelity parameter proposed by \cite{Rybizki2022}, namely $fidelity\_v2$, which classifies {\it Gaia} sources using neural network models trained on a dataset of \emph{good} and \emph{bad} sources. For this parameter, we adopted a minimum threshold of 0.5. Parallaxes zero-point values were derived following the \cite{Lindegren2021parallax} recipe\footnote{\href{https://pypi.org/project/gaiadr3-zeropoint/}{https://pypi.org/project/gaiadr3-zeropoint/}}.

For each of the $\sim$ 190 million stars meeting our quality criteria, we derived 1,000 estimates of the total velocity ($v_{\rm T}$) in the Galactic Standard of Rest frame, using a multivariate Gaussian distribution. This distribution accounts for the parameters provided by Gaia and their associated uncertainties. The correlations between parallax and proper motion were considered through the covariance matrix detailed in the Appendix, while uncertainties related to sky positions were assumed to be negligible. For each iteration, we estimated the hypothetical heliocentric radial velocity ($v_{\rm los_{\rm 0}}$) that minimizes $v_{\rm T}$, which varies with the position within the Galaxy (e.g. $v_{\rm los_{\rm 0}}$ = 0\,km\,s$^{-1}$ for stars around $\ell$ = 180$^{\circ}$). This means that, although we refer to these values as total velocities, they are actually minimum total velocities. Thus, this approach provides a conservative method for identifying HVS candidates when assuming $v_{\rm los_{\rm 0}}$ to derive their total velocities.

Throughout this calculation, the distance from the Sun to the Galactic center and to the plane are set, respectively, at 8.2 and 0.025\,kpc \citep{Bland-Hawthorn2016}. The solar velocity vector is set at $v_{\odot} = [11.1,232.2,7.25]$\,km\,s$^{-1}$ \citep{Schonrich2010,Bovy2012}. For each of the iterations (i.e. positions within the Galaxy), we derived the $v_{esc}$ in the MWPotential2014 potential provided by \textsc{GALPY} \citep[see description in ][]{Bovy2015} and assumed that $v_{esc}$ is independent of $z$ (our values at any $R$ correspond to $z = 0$\,kpc). We verified that for most of the sources, the dispersion in $v_{esc}$ remained below 2\,km\,s$^{-1}$. Of course, the  $v_{esc}$ values depend on the Galactic potential selected, although we assumed that this effect is negligible in a first approach (see discussion in Section \ref{potentials}). 

\begin{figure}
    \centering
    \includegraphics[width=\columnwidth]{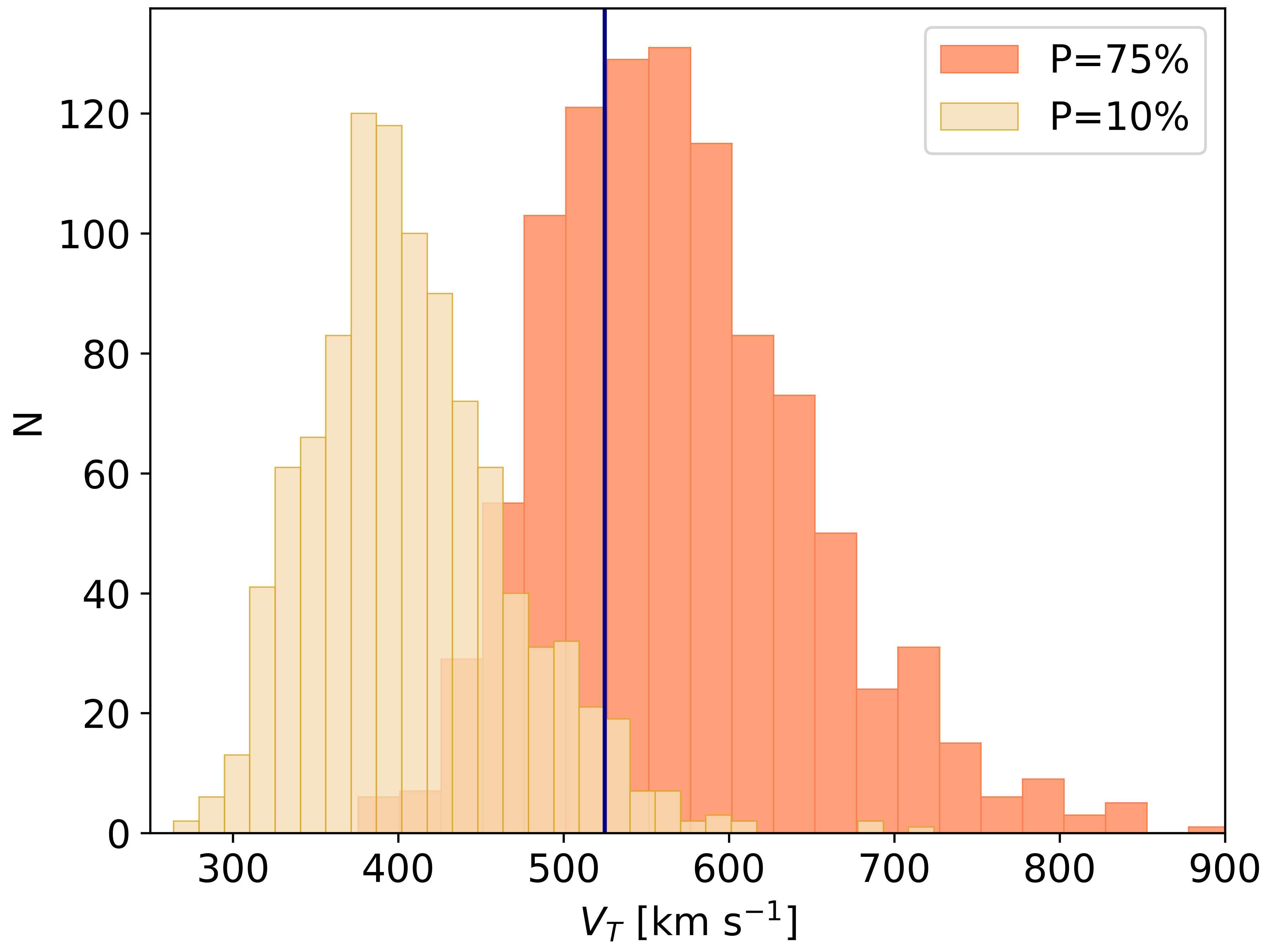}
    \caption{$v_{\rm T}$ distribution from 1,000 realizations for two HVS candidates with similar median $v_{\rm esc}$ ($\sim 500$\,km\,s$^{-1}$; blue vertical line). The yellow and orange distributions are associated with stars with initial escape probabilities of $P_{\rm esc, 0} = $ 10$\%$ and 75$\%$, respectively.}
    \label{fig:plot1}
\end{figure}

The fraction of realizations where $v_{\rm T} > v_{esc}$ is considered an estimate of the probability ($P_{\rm esc}$) of being a HVS. Since it is only possible to compute the real probability when $v_{\rm los}$ is known, we denote this initial estimate as $P_{\rm esc, 0}$. Figure \ref{fig:plot1} illustrates how the probability is computed for two HVS candidates with a similar median $v_{esc}$ $\sim$ 500\,km\,s$^{-1}$. While for one of the stars, the $v_{\rm T}$ distribution lies almost completely below $v_{esc}$ ($P_{\rm esc, 0}= 10\%$), the second $v_{\rm T}$ distribution is associated with a star with $P_{\rm esc, 0} \sim 75\%$. In both cases, it would only be possible to confirm their nature once we obtain reliable $v_{\rm los}$ and derive their 3D velocity vectors. 

\begin{figure*}
    \centering
    \includegraphics[width=\textwidth]{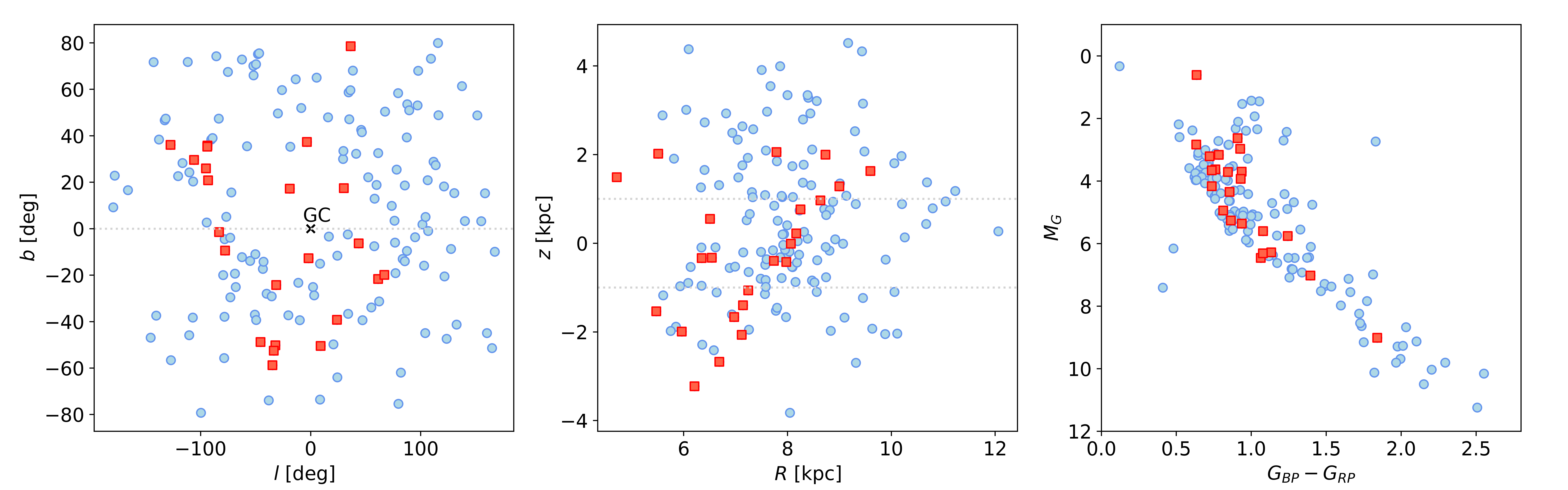}
    \caption{Spatial distribution of the HVS candidates in the $\ell - b$ and $Y - X$ planes, in the left and central panels, respectively. Objects for which we obtained follow-up spectroscopy are marked with red squares. Right: position of the HVS candidates and targets observed in the color-magnitude diagram (absolute magnitudes).}
    \label{fig:plot2}
\end{figure*}

Our final sample of candidates, with $P_{\rm esc, 0} \ge 50\%$, contains 149 sources with the Galactic distribution shown in Figure \ref{fig:plot2}. As expected, the basic quality cuts applied to Gaia data restrict our sample to nearby objects ($\sim 40\%$ within 2\,kpc from the Sun), and $\sim 70\%$ of them are immersed in the Galactic disk assuming a height scale of 1\,kpc \citep[e.g.][]{Bland-Hawthorn2016}.  However, despite the quality criteria applied, it is still possible that a few spurious sources remain in our sample. To further assess astrometric reliability, we followed the additional recommendations by \citet{Scholz2024} for identifying HVS candidates with the most robust solutions. We cross-matched our sources between {\it Gaia} DR2 and DR3 to identify any significant discrepancies in proper motions and parallaxes. Additional quality checks included verifying the absence of nearby neighbors, ensuring a high number of individual observations, and confirming the lack of astrometric warning flags in {\it Gaia} DR3.

Out of the 149 sources, 36 did not pass these final quality criteria. We therefore classify them as part of a {\emph{silver}} subsample, whereas the remaining 113 stars constitute the astrometrically reliable {\emph{golden}} subsample. In this work, all targets are analyzed regardless of this classification, but the distinction is important when evaluating the robustness of the HVS nature of each candidate. Even if not all stars in the golden sample turn out to be true HVSs, the number of candidates remains significant given the stringent quality cuts applied during the selection.

\begin{table*}
    \centering
    \caption{Sample of HVS candidates selected for follow-spectroscopy. }
    \begin{tabular}{lclllclc}
   \hline
   ID  & Class. & {\it Gaia} DR3 source ID &  R.A. &  Dec. & P$_{\rm esc, 0}$ & t$_{\rm exp}$ & G  \\
       &          & & (J2000.0) & (J2000.0) & (\%) &  (s)   & (mag)\\
   \hline
HVS01  & S & 3841458366321558656  & 09:34:09.21 & +01:35:16.7 & 100 & 900 (1) & 15.9 \\
HVS02  & G & 1451652599056932480  & 13:43:14.20 & +27:19:53.2 & 91.4 & 2400 (2) & 17.2 \\
HVS03 & G & 4294774301679435776 & 19:35:58.47 & +06:25:17.3 & 84.3 &  600      & 16.0 \\
HVS04  & G & 5248871805803805440  & 09:22:05.82 & -64:01:23.1 & 81.2 &  900 (1) & 16.0 \\
HVS05 & G & 6496426077183248256 & 23:30:49.78 & -54:49:57.5 & 80.4 & 1500      & 17.3 \\
HVS06  & G & 5673818825000094336  & 09:58:09.80 & -16:41:14.8 & 77.3 &  900 (1) & 16.2 \\
HVS07 & G & 6389682292902592256 & 23:33:07.86 & -66:24:26.6 & 76.5 &  900      & 16.4 \\
HVS08 & G & 6733389486012783488 & 18:33:00.31 & -36:24:22.8 & 73.2 & 2400 (2) & 17.3 \\
HVS09 & G & 6587991790636824960 & 21:49:19.10 & -35:22:18.3 & 67.7 & 2820      & 17.2 \\
HVS10  & G & 3555961002414624640  & 10:49:01.94 & -18:05:19.7 & 64.5 & 2100 (2) & 16.8 \\
HVS11 & G & 6407392126691609600 & 22:39:31.11 & -59:58:10.7 & 62.5 & 1800      & 16.6 \\
HVS12 & S & 6318336357766371072 & 15:31:46.17 & -08:13:12.6 & 59.1 &  600      & 15.8 \\
HVS13 & S & 6015520135250185088 & 15:48:55.11 & -32:24:09.8 & 57.9 & 2400      & 17.0 \\
HVS14 & G & 6491701097761578240 & 22:57:56.68 & -58:59:04.3 & 56.1 &  600      & 15.9 \\
HVS15 & G & 6807757433152562048 & 21:05:39.59 & -23:09:16.7 & 54.7 & 1200      & 16.2 \\
HVS16 & G & 6434875794219996928 & 18:45:46.80 & -66:35:59.8 & 54.1 &  600      & 16.1 \\
HVS17 & G & 1758246228142845696 & 21:03:55.53 & +12:58:19.8 & 53.2 &  600      & 15.9 \\
\hline
HVS18  & S & 5460314969926337024  & 10:11:05.07 & -30:55:47.9 & 46.4 & 1800 (2) & 16.5 \\
HVS19  & S & 3555997801694385664  & 10:48:56.48 & -17:58:47.6 & 43.0 &  900 (1) & 15.3 \\
HVS20 & S & 4473731089063378176 & 17:43:44.52 & +05:31:54.1 & 37.7 &  600      & 14.1 \\
HVS21 & S & 1788329175516075264 & 21:09:32.38 & +18:34:41.7 & 37.2 &  600      & 15.8 \\
HVS22 & S & 5308961456302889472  & 09:35:26.18 & -54:17:29.5 & 17.4 & 2400 (2) & 17.2 \\
HVS23  & S & 5472005458587726976  & 10:18:04.17 & -26:07:25.3 & 13.6 & 1800 (2) & 17.4 \\
    \end{tabular}
    \label{table_fors}
     \tablefoot{Columns list: Star ID, Classification (Golden/Silver), {\it Gaia} DR3 source ID, position, $P_{\rm esc, 0}$, exposure times, and $G$ magnitudes for the HVS candidates selected for FORS2 follow-up spectroscopy. The stars below the horizontal line have $P_{\rm esc, 0} < 50\%$. The numbers in parentheses indicate the number of exposures.}
\end{table*}

\section{Spectroscopic follow-up with FORS2}

We have focused on HVS candidates that lack {\it Gaia}-provided $v_{\rm los}$ estimates. However, even if their transverse velocities ($v_{\rm t}$\footnote{$v_{\rm t} = (4.74/\varpi)\sqrt{\mu_\alpha^2 + \mu_\delta^2}$}) suggest a total Galactocentric velocity above the local $v_{\rm esc}$ value, the origin of the stars cannot be established without full 6D phase-space information. Therefore, we continued our study by deriving $v_{\rm los}$ from follow-up spectroscopy for a subset of candidates. 

 A selection of relatively bright HVS candidates from both the golden and silver subsamples, spanning a wide range of $P_{\rm esc, 0}$ values and observable over two semesters, was chosen for spectroscopic follow-up (see Table \ref{table_fors}). Three of them, namely HVS01, HVS02, and HVS03, have been previously proposed as HVSs based on their $v_{\rm t}$ only \citep{Du2019,Scholz2024}, so we included them in our sample to derive their $v_{\rm los}$ and investigate their origin within the Galaxy. Six targets with $P_{\rm esc, 0}$ below 50\% were wrongly included in the preliminary list ($P_{\rm esc, 0}$ values changed after a modification of our HVS identification procedure), but we used them here as control stars\footnote{At the time of the first observing run, we had assumed $v_{\rm los_{\rm 0}}$ = 0\,km\,s$^{-1}$ for all the stars, independently of its position within the Galaxy.}. 
 
 It is important to note that HVS01 was excluded by \citep{Scholz2024} and has been classified as a carbon-enhanced metal-poor (CEMP) star candidate \citep{Li2018,Fang2025}. Given that certain types of CEMP-s stars are known to exhibit radial velocity variability as a result of binarity \citep[e.g.][]{Placco2014,Starkenburg2014}, caution is advised when interpreting the properties of this object. Nevertheless, we retain HVS01 in our sample due to the relevance of such peculiar stars for studies in Galactic archaeology. As for HVS20, it is classified as a c-type RR\,Lyrae by {\it Gaia} \citep{Clementini2023,Eyer2023}, with an amplitude of $\sim 0.4$\,mag in $G - $band and a period of $\sim 7$\,h. The variability of HVS20 might impact the determination of $v_{\rm los}$ and introduce an error of up to $\sim$ 25\,km\,s$^{-1}$  \citep[see][and references therein]{Prudil2024}. 

Sources had magnitudes in the range $14.1 < G < 17.4$ (see their spatial distribution and their positions in the color-magnitude diagram in Figure \ref{fig:plot2}), and were observed with FORS2 from the Paranal Observatory as part of an "all-weather" filler program (111.24P0.006, 110.23UC.007). A total of 23 stars were observed, accounting for a total time of 21.6\,h. The observations were acquired in Service Mode, in the long-slit mode, using the grism 1200R+93 and 1.0\,arcsec slit ($\mathcal{R} \sim$ 2140). The exposure times ranged from 600\,s to $2 \times 2400$\,s (see Table \ref{table_fors}). The data were reduced with the FORS2 v5.6.4 reduction pipeline, which included flat fielding, wavelength calibration, correction of spatial distortion, sky subtraction, optimal extraction of spectra, and flux calibration. All spectra were corrected for the heliocentric velocity using \verb|MOLLY|\footnote{\url{https://cygnus.astro.warwick.ac.uk/phsaap/software/molly/html/INDEX.html}}. To ensure accurate wavelength calibration, the spectra were subsequently aligned by referencing the position of the sky emission line at 6300.34\,\AA. The same emission line was re-identified in the corrected spectra, yielding a mean residual for the centroid of the line better than 5\% of the spectral dispersion across the entire sample.

$v_{\rm los}$ and a rough estimate of $[{\rm Fe/H}]$ were derived for each star using \textsc{Doppler}.\footnote{https://github.com/dnidever/doppler}, which is designed to characterize stars by convolving a model spectrum to the resolution or Line Spread Function (LSF) of the observed spectrum. Uncertainties associated with the parameters are the $1\sigma$ of the distributions obtained by bootstrapping 10,000 times the observed spectra. In the same way, a parallel approach with {\tt pySME} \citep{pysme} was performed to derive $v_{\rm los}$ and $[{\rm Fe/H}]$. We used typical values for micro and macroturbulence for dwarf stars, while using a fixed photometric log$g$ during the fitting process to reduce the number of free parameters. The values are in agreement with {\tt Doppler} and consistent for main-sequence stars (see right panel in Figure \ref{fig:plot2}); however, the low spectral resolution results in considerable uncertainties.

\section{Results and discussion}

\begin{table*}
\centering
\caption{Kinematic properties and classification of the analyzed HVS candidates.}
\begin{tabular}{lllllllll}
\hline
 ID & $P_{\rm esc}$ & $v_{\rm los}$ & $v_{\rm T}$ & $z$ & $r_{\rm min}$ & $t_{\rm rmin}$ & $t_{\rm z=0}$ & Classification\\
    & (\%) & (km/s) & (km/s) & (kpc) & (kpc) & (Myr) & (Myr) &  \\
\hline
HVS01 & 100.0 & -100.3 $\pm$ 8.8 & 900.8 $\pm$ 115.6 & 1.7 $\pm$ 0.2 & - & - & - & Extra-Galactic\\
HVS02 & 92.7 & -155.3 $\pm$ 5.1 & 680.1 $\pm$ 140.7 & 2.1 $\pm$ 0.4 & 8.0 $\pm$ 0.3 & -2.0 $\pm$ 0.7 & -83.1 $\pm$ 30.4 & Disk crossing / Extra-Galactic\\
HVS03 & 86.8 & -247.1 $\pm$ 6.0 & 628.1 $\pm$ 97.6 & -0.3 $\pm$ 0.1 & - & - & - & In disk. No crossing\\
HVS04 & 93.7 & 401.2 $\pm$ 6.2 & 580.4 $\pm$ 45.6 & -0.4 $\pm$ 0.1 & 3.5 $\pm$ 0.3 & -12.2 $\pm$ 0.9 & -0.3 $\pm$ 0.1 & Disk crossing\\
HVS05 & 83.7 & 82.8 $\pm$ 7.0 & 611.0 $\pm$ 125.5 & -2.1 $\pm$ 0.4 & - & - & - & Extra-Galactic\\
HVS06 & 78.9 & 234.0 $\pm$ 7.3 & 549.7 $\pm$ 71.2 & 1.3 $\pm$ 0.2 & 4.4 $\pm$ 0.7 & -14.4 $\pm$ 2.1 & - & Extra-Galactic\\
HVS07 & 78.2 & 183.5 $\pm$ 6.1 & 591.8 $\pm$ 92.9 & -2.8 $\pm$ 0.4 & 1.4 $\pm$ 0.4 & -10.8 $\pm$ 1.5 & -9.0 $\pm$ 0.5 & Disk crossing. Hills?\\
HVS08 & 76.0 & 13.0 $\pm$ 4.3 & 599.3 $\pm$ 140.5 & -0.3 $\pm$ 0.1 & - & - & - & In disk. No crossing\\
HVS09 & 73.2 & -138.0 $\pm$ 5.3 & 565.8 $\pm$ 73.9 & -1.1 $\pm$ 0.1 & 3.2 $\pm$ 0.5 & -11.3 $\pm$ 1.9 & -3.0 $\pm$ 0.1 & Disk crossing\\
HVS10 & 74.5 & 327.3 $\pm$ 5.7 & 550.1 $\pm$ 57.8 & 0.8 $\pm$ 0.1 & - & - & - & In disk. No crossing\\
HVS11 & 77.5 & -96.4 $\pm$ 7.2 & 581.6 $\pm$ 80.2 & -1.7 $\pm$ 0.2 & 7.2 $\pm$ 0.3 & -0.5 $\pm$ 0.2 & - & Extra-Galactic\\
HVS12 & 61.1 & -55.9 $\pm$ 4.5 & 563.7 $\pm$ 105.9 & 2.1 $\pm$ 0.3 & - & - & - & Extra-Galactic\\
HVS13 & 60.1 & 72.1 $\pm$ 5.2 & 553.5 $\pm$ 112.2 & 0.6 $\pm$ 0.1 & 5.7 $\pm$ 0.4 & -6.1 $\pm$ 1.7 & -5.5 $\pm$ 51.2 & Disk crossing\\
HVS14 & 59.2 & 59.6 $\pm$ 6.1 & 556.5 $\pm$ 105.8 & -3.3 $\pm$ 0.5 & 6.8 $\pm$ 0.3 & -3.3 $\pm$ 0.9 & - & Extra-Galactic\\
HVS15 & 57.8 & -29.5 $\pm$ 8.5 & 554.8 $\pm$ 120.7 & -2.0 $\pm$ 0.4 & - & - & -21.1 $\pm$ 0.8 & Disk crossing\\
HVS16 & 56.0 & 66.1 $\pm$ 8.3 & 557.7 $\pm$ 81.3 & -1.6 $\pm$ 0.3 & - & - & - & Extra-Galactic\\
HVS17 & 57.3 & -259.5 $\pm$ 5.7 & 535.2 $\pm$ 99.4 & -1.5 $\pm$ 0.3 & - & - & - & Extra-Galactic\\
HVS18 & 49.2 & 267.8 $\pm$ 7.7 & 502.6 $\pm$ 70.1 & 1.0 $\pm$ 0.1 & 2.7 $\pm$ 0.7 & -15.5 $\pm$ 1.9 & -35.9 $\pm$ 8.3 & Disk crossing\\
HVS19 & 43.3 & 214.6 $\pm$ 6.1 & 500.4 $\pm$ 43.1 & 2.0 $\pm$ 0.2 & - & - & -6.1 $\pm$ 0.4 & Disk crossing\\
HVS20 & 38.2 & -108.6 $\pm$ 9.1 & 550.4 $\pm$ 59.3 & 1.5 $\pm$ 0.1 & - & - & -2.3 $\pm$ 0.1 & Disk crossing\\
HVS21 & 38.7 & -209.5 $\pm$ 9.3 & 517.4 $\pm$ 28.2 & -0.4 $\pm$ 0.1 & - & - & - & In disk. No crossing\\
HVS22 & 19.5 & 216.5 $\pm$ 7.1 & 494.7 $\pm$ 27.0 & -0.1 $\pm$ 0.1 & - & - & - & In disk. No crossing\\
HVS23 & 16.9 & 274.8 $\pm$ 2.3 & 504.8 $\pm$ 20.8 & 0.2 $\pm$ 0.1 & - & - & - & In disk. No crossing\\
\hline
\end{tabular}
\tablefoot{ Columns list: (1) Star ID, (2) probability of being unbound from the Milky Way ($P_{\rm esc}$), (3) radial velocity ($v_{\rm los}$), (4) total Galactocentric velocity ($v_{\rm T}$), (5) vertical position relative to the Galactic plane ($z$), (6) minimum Galactocentric distance reached in the past ($r_{\rm min}$), (7) time when star reached its minimum Galactocentric distance ($t_{\rm rmin}$), (8) time when star crossed the Galactic plane ($t_{\rm z=0}$), and (9) classification based on orbital properties. The classification labels indicate whether the stars are of hypothetical extra-Galactic origin, cross the Galactic plane, or remain within the disk without crossing it.}
\label{table_results}
\end{table*}

\begin{figure}
    \centering
    \includegraphics[width=\columnwidth]{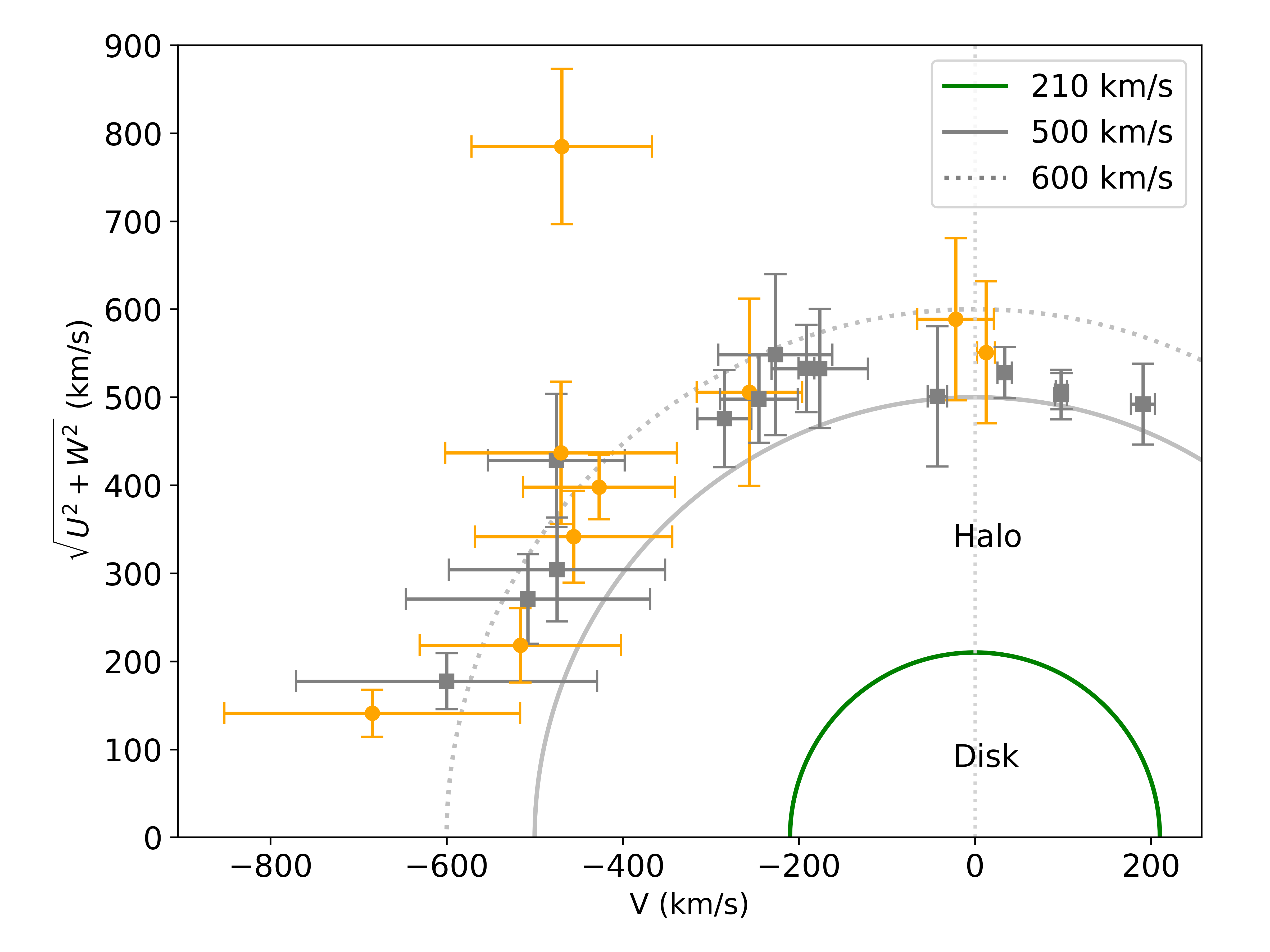}
    \caption{Toomre diagram showing the kinematic distribution of our stars in the (V, $\sqrt{U^2 + W^2}$) plane. The green solid curve represents a total space velocity of 210\,km\,s$^{-1}$, approximately delimiting the thick disk and halo populations \citep{Gaia_cartography}, while the solid and dotted grey curves mark the 500 and 600  km\,s$^{-1}$ contours, respectively. Objects highlighted in orange correspond to those identified as having an extra-Galactic origin in this work. The target with maximum $\sqrt{U^2 + W^2}$ value is HVS01.} 
    \label{fig:plot3}
\end{figure}

The probability of being an HVS ($P_{\rm esc}$), computed using the $v_{\rm los}$ derived from FORS2 data, is in good agreement with the values previously obtained ($P_{\rm esc, 0}$), as calculated in the absence of $v_{\rm los}$ measurements. Since the $v_{\rm T}$ distributions previously derived above represent lower limits, $P_{\rm esc}$ > $P_{\rm esc, 0}$. Therefore, our HVS candidates retain this classification once new information is incorporated. Although a population of HVSs is not expected to follow an isotropic velocity distribution, the fact that $v_{\rm T}$ significantly exceeds $\sqrt{2}\,v_{\rm t}$ for most of our targets may indicate potentially overestimated proper motion values \citep[see also][]{Palladino2014, Scholz2024}. While our sources are not necessarily affected by poor astrometry, this observation warrants caution. The kinematics of these targets should be checked once future {\it Gaia} data releases become available. Nonetheless, as shown in the Toomre diagram (Figure \ref{fig:plot3}), all of our targets exhibit halo-like kinematics, which is expected for such fast-moving objects: in the (V, $\sqrt{U^2 + W^2}$) plane, all the stars lie beyond the 500\,km\,s$^{-1}$ contour.

To investigate the origin of our targets, we computed their mean orbital trajectories over the past 1\,Gyr, based on 1000 Monte Carlo realizations with a time step of 1\,Myr, adopting the \texttt{MWPotential2014} Galactic potential. We first verify whether any of our HVS candidates may originate via the Hills mechanism \citep{Hills1988}, assuming that this will be the case of objects whose minimum Galactocentric distance along its orbit is $r < 1$\,kpc. This is similar to the criteria used in previous works \citep[e.g.][]{Marchetti2019,Koposov2020,Liao2023}. In our sample, none of them satisfy this condition with the exception of HVS07, which has a past orbit that approximates to the inner MW at  $r \sim 1.4 \pm 0.4$\,kpc from its center, almost 11\,Myr ago, when $v_{\rm T}$ = 710 $\pm$ 70\,km\,s$^{-1}$. This velocity lies near the lower limit of the range of ejection velocities predicted for intermediate mass stars produced via the Hills mechanism in the vicinity of the Galactic SMBH \citep[e.g.][]{Bromley2006}. While such low velocities are not excluded - particularly for intermediate/low-mass stars, which can remain bound - this nonetheless casts doubt on a Galactic Center origin for this HVS candidate.

Although our results may depend on the assumed Galactic potential and be influenced by astrometric uncertainties, they remain consistent with previous systematic studies, which found little evidence supporting a Galactic center origin for HVSs \citep[e.g.][]{Kreuzer2020,Irrgang2021}. Indeed, despite the hundreds of confirmed (and refuted) HVSs, only a few dozens seem to have their origin ``close'' to Sgr A$^{*}$, being S5-HVS1 one of the most promising candidates \citep{Koposov2020}. Therefore, even if the sample presented in this work is limited, it is clear that the family of Galactic HVSs must have been ejected from other locations within the MW, not necessarily its center.   

\begin{figure}
    \centering
    \includegraphics[width=\columnwidth]{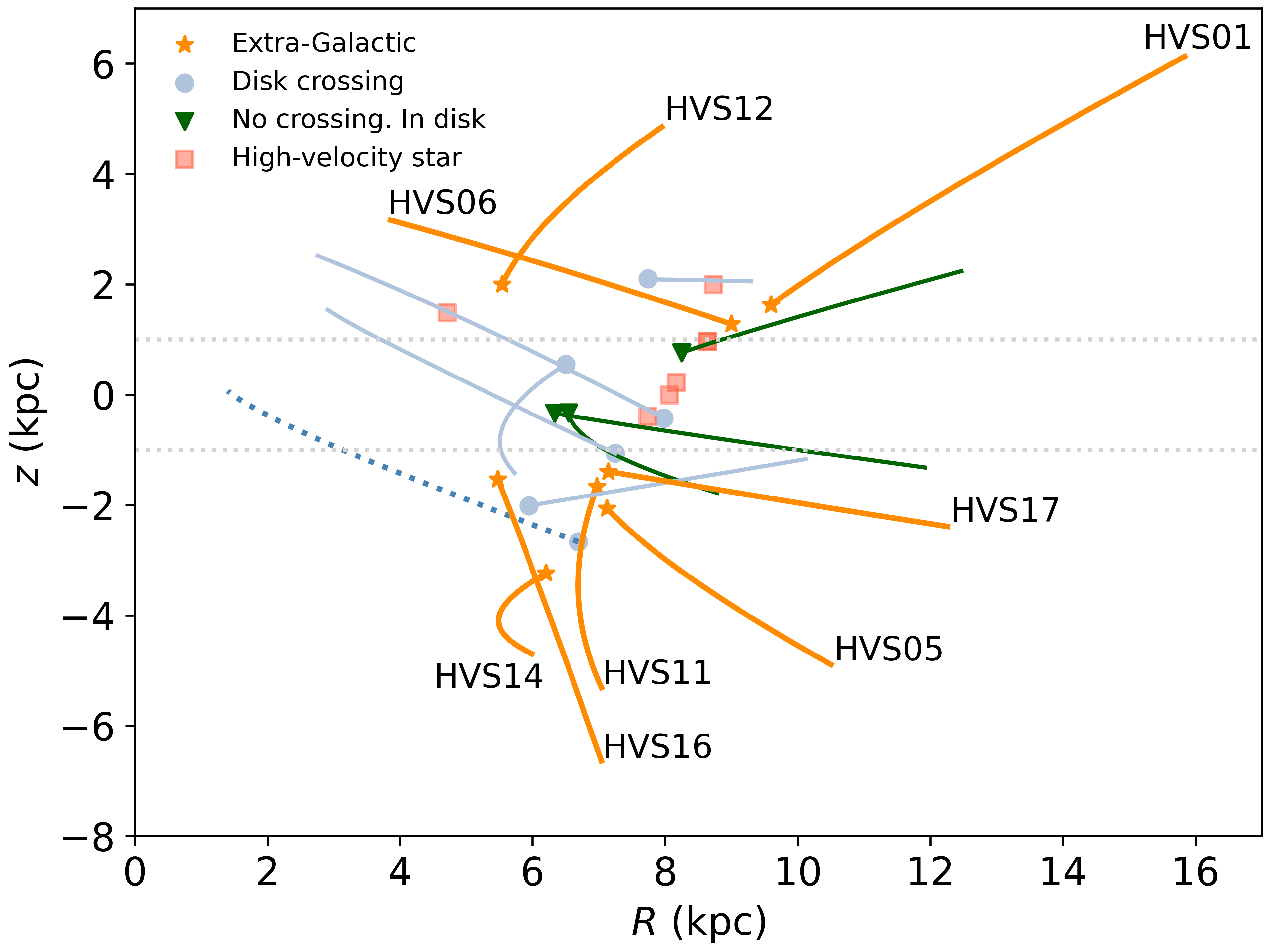}
    \caption{Past orbits computed using \textsc{GALPY} (showing here only the last 10\,Myr), based on {\it Gaia}+FORS2 data. The orange, blue, and green solid lines represent different orbital classifications: orange corresponds to targets with a hypothetical extra-Galactic origin, blue indicates orbits that cross the Galactic plane ($b = 0^{\circ}$), and green represents orbits of objects currently confined within the Galactic disk which have not crossed the plane in the past. The dashed blue line corresponds to HVS07, the only HVS candidate which approaches the Galactic center. Grey squares indicate the position of the non-HVS sources.}  
    \label{fig:plot4}
\end{figure}

We classify our targets considering not only their trajectories but also their current positions within the Galaxy. We will group the targets in 4 categories, similar to the scheme proposed by \cite{Marchetti2019}: {\it i)} HVSs which are not in the Galactic disk and whose past orbits do not intersect the plane, which we classify as extra-Galactic star candidates (although halo stars in extremely eccentric orbits may fall into this category); {\it ii)} HVSs whose past orbits do not intersect the plane ($b = 0^{\circ}$) but are currently located within the Galactic disk ($|z| \leq 1$\,kpc); {\it iii)} HVSs whose past orbits intersect the MW disk; and {\it iv)} stars with low probability of being a HVS, but still are HiVel stars. With these basic definitions, we assume that HVSs with past orbits crossing the Galactic disk might be originated in this dense MW component, although it is not possible to completely rule out their extra-Galactic origin. The suggested classification is also included in Table \ref{table_results}, while their recent past orbits and current positions within the Galaxy are shown in Figure \ref{fig:plot4}. We include in the table under the category “disk crossing” only those stars whose past orbits intersect the Galactic plane in more than half of the realizations.

The fastest star in our sample (HV01; $v_{\rm T}\sim 900$\,km\,s$^{-1}$) and other 7 targets have past orbits throughout the Galaxy that do not intersect the Galactic plane during the last 1\,Gyr. The only exception is HVS02, which crossed the Galactic disk approximately 8\,Myr ago, but at a Galactocentric distance of $R \sim 50$\,kpc. Although the exact extent of the Galactic disk remains uncertain \citep[e.g.][]{Lian2024}, several HVSs have been discovered in the outer disk at $R > 20$\,kpc, suggesting a possible connection with the impact of satellite galaxies \citep{Irrgang2021}. We therefore also classify HVS02 as a tentative extra-Galactic star.

All these stars, except HVS06, are on retrograde orbits around the Galaxy (see Figure \ref{fig:plot3}), which has been considered an indicative (in combination with other indicators) of the possible extra-Galactic origin of globular clusters and stars within the MW \citep[e.g.][]{Koppelman2019,Matsuno2019,Myeong2019}. Such an observation strengthens our classification; however, they might still be in-situ formed halo stars and the role of the bars on producing stars on retrograde orbits remains unclear \citep{Fiteni2021}. In a Galactic halo built through the accretion of numerous protogalactic fragments \citep[e.g.][]{Mackereth2019,Helmi2020}, it remains very plausible that most fast-moving halo stars are associated with past merger events in the MW.

We have identified a second group of stars which do not approach the MW center and whose past orbits do not intersect the Galactic plane ($b = 0^{\circ}$) during the last 1\,Gyr. However, unlike the previous subgroup, these stars are currently immersed in the Galactic disk ($|z| \leq 1$\,kpc). Although it is still possible that these stars were originated far away from the disk, we cannot rule completely out the possibility of being generated by other mechanisms within the MW plane. From our targets, seven of them fall in this category, but only three of them - HVS03, HVS08, and HVS10 - are confirmed as HVSs. The rest of stars in this category are HiVel stars with $P_{\rm esc} \leq 50\%$ and  $v_{\rm T} \sim 500$\,km\,s$^{-1}$, which are still remarkable velocities in the Galactic context.  

The remaining stars in our sample have past orbits that intersect the Galactic plane, with three of them exhibiting $P_{\rm esc} \leq 50\%$. It is also likely that these HVSs crossing the disk have an extra-Galactic origin, but the available data are insufficient to confirm that hypothesis. 

While a precise chemical characterization of these targets is beyond the scope of this work, we derived approximate $[{\rm Fe/H}]$ estimates using the \textsc{Doppler} code. These values should be considered indicative only, due to the limited spectral resolution and the reduced sensitivity of the spectral templates in the metal-poor regime. Roughly speaking, the sample has a mean metallicity of $\langle [{\rm Fe/H}] \rangle \sim -1.9$, with 13 stars having $[{\rm Fe/H}] \leq -2$, and three stars with $[{\rm Fe/H}] > -1$. The group of stars classified as “Extra-Galactic” exhibits very low metallicities, clustering around $[{\rm Fe/H}] \sim -2.5$, a value likely reflecting the lower boundary of the spectral template’s applicability rather than a physically meaningful estimate. Nevertheless, these stars constitute high-priority targets for follow-up high-resolution spectroscopy aimed at constraining their origins through chemical tagging. Interestingly, HVS07 - our only target with a past trajectory marginally compatible with an origin near the Galactic center - also belongs to this extremely metal-poor group. This further argues against the classical Hills mechanism (binary disruption near the SMBH) as the dominant ejection process for this object.

On the other hand, the apparently most metal-rich star in the sample, HVS16, does not intersect the Galactic plane in its orbital history. Stars with such metallicities and velocities exceeding the local escape speed are promising candidates for an origin in MW satellites such as the LMC or the Sagittarius dwarf galaxy. Finally, stars that crossed the Galactic disk within the last 1\,Gyr show a broader metallicity range, suggesting a diversity of stellar populations and possible origins.

\subsection{Other possible origins for the HVSs}

Although securely pinpointing the place of origin of these stars requires not only computing their orbits but also incorporating additional information (e.g., chemical abundances), we proceed by exploring some of the hypotheses proposed in the literature and examining the spatial coincidence of our candidates with different systems/objects in the MW.

The center of the Galaxy is not the only place where binary systems can interact with black holes. Less massive black holes, such as the elusive IMBHs, are theoretically capable of disrupting binary systems and generating HiVel stars or even HVSs \citep{Fragione2016,Fragione2019}. Evidence of the presence of IMBHs in Galactic globular clusters has been gathered during recent years, despite the observational effort required to unveil those objects in dense environments \citep[e.g.][]{Giesers2018}. Recently, \cite{Haberle2024} found evidence of the existence of fast-moving stars with total velocities above the local $v_{\rm esc}$ (cluster), although the number of these objects observed is too high to be explained only through the binary-IMBH interaction mechanism. Nevertheless, different simulations have shown that this type of interaction in globular clusters could generate stars traveling at velocities as fast as $\sim$2000\,km\,/s \citep{Cabrera2023}.

We investigated the potential ejection of HVSs from Galactic globular clusters by reconstructing their past orbits. This analysis used the potential model MWPotential2014, along with the {\it Gaia}\,DR3 astrometric data for the globular clusters \citep{Baumgardt2021,Vasiliev2021}. For this purpose, we computed 1,000 orbital trajectories for both the target stars and the globular clusters and analyzed the resulting minimum relative distance distributions. As a selection criterion, stars are considered potentially associated with a given cluster only if their mean past orbits approach within a minimum relative distance of 500\,pc from the mean orbit of the cluster. Although the uncertainties in the astrometric parameters make these associations speculative, four stars satisfy the aforementioned criterion: HVS04, HVS07, HVS09, and HVS13. These stars approached NGC\,6171, Terzan\,6, IC\,1276, and NGC\,6121  at relative distances of $d$ = 0.3$\pm$0.5 (10\,Myr ago), 0.4$\pm$ 0.4 (10\,Myr), 0.4$\pm$0.7\,kpc (5\,Myr), and 0.1$\pm$ 0.1\,kpc (1\,Myr), respectively. However, if we further restrict the minimum relative distance to $d_{\rm min}$ = 100\,pc and compute the probability of having a relative distance below that threshold, we find that only HVS13 might be associated with NGC\,6121 with $P_{\rm cl}$ = 36$\%$. Consequently, these associations, although possible, cannot be confirmed given the current level of uncertainty.

The destruction of binary systems after a supernova detonation is also among the plausible scenarios of HVSs formation \citep[see][and references therein]{Ruiz-Lapuente2023}. In this case, the association of a HVS with a supernova remnant (SNR) is hindered by the difficulty to derive accurate distances and proper motions for the latter. In this case, we simply searched for HVSs whose past orbits intersect with the current sky position of known SNRs from the \cite{Ferrand2012} catalog\footnote{http://snrcat.physics.umanitoba.ca}. Only coincidences with a projected minimum distance of $\leq$ 10\,arcmin during the last 5\,Myr  are considered. HVS09, HVS13, and HVS20 (all of them crossed the disk) might be associated with G018.0-00.7, G018.5-00.4, and G036.6+02.6, respectively. All these crossings took place between 2.5 and 3.9\,Myr ago but, unfortunately, the mentioned SNRs are not completely characterized (e.g. distances). SNRs lifetimes are of around 10$^6$\,yr under favorable environmental conditions \cite[see][]{Bamba2022}, so our larger hypothetical ejection times would discard such associations. 

MW satellite galaxies might host massive black holes (MBHs), which could generate mechanisms similar to the Hills mechanism. These processes would provide nearby stars with enough energy to escape their host galaxies. For instance, \cite{Li2022} identified 60 HiVel stars probably originated in the Sagittarius dwarf spheroidal galaxy, with at least 2 of them classified as HVSs. The LMC is also a well-established generator of HVSs, with a few examples of HVSs with reconstructed orbits pointing towards its past position \citep[e.g.][]{Edelmann2005,Irrgang2018,Erkal2019}, implying the presence of MBHs \citep[see also][]{Han2025}. We have repeated the same procedure followed for the GCs described above and assuming proper motions, distances, and $v_{\rm los}$ from the literature for LMC \citep{Pietrzynski2019,McConnachie2012} and Sagittarius \citep{Gaia2018sat}. None of the HVSs included in this work approached in the past to those systems close enough to suggest an association: the closest encounter between the center of the LMC and one of our target stars have a minimum relative distance of $d = 18 \pm 25$\,kpc (HVS20) taking place $\sim 230$\,Myr ago; as for Sagittarius, the star with closest past trajectory is also HVS20, with a minimum relative distance of $d = 10 \pm 3$\,kpc. The association of HVS20 - an RR-Lyrae variable - with the LMC is compatible within the current uncertainties. However, due to its pulsating nature, the $v_{\rm los}$  of HVS20 is likely to be poorly constrained from a single-epoch spectrum. A dedicated spectroscopic follow-up covering multiple phases of its pulsation cycle would be essential to better constrain its systemic radial velocity and test this intriguing association.

In conclusion, none of the HVS candidates included in this work presents past orbits clearly compatible with that of the MW satellites analyzed here or the current position of known Galactic SNRs. Future {\it Gaia} data releases with more accurate astrometry, and the obtaining of higher-precision velocities, would allow us to better establish the probable crossing area through the Galactic plane, and reconstruct the past orbits of HVSs in comparison with those of other progenitor stellar systems.

\begin{figure}
    \centering
    \includegraphics[width=\columnwidth]{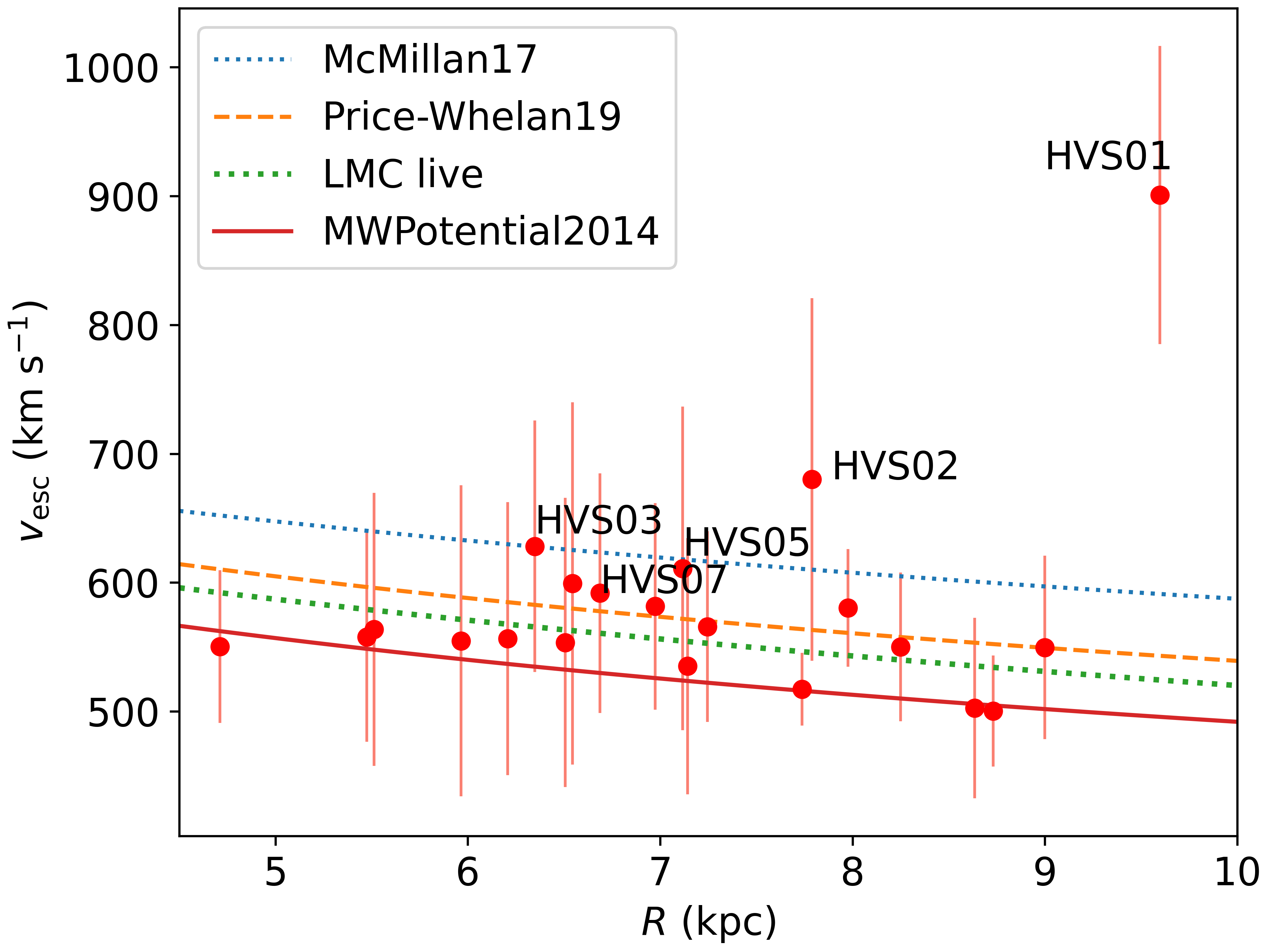}
    \caption{$v_{\rm esc}$ as a function of $R$ for different Galactic potentials. The red solid line corresponds to the MWPotential2014 from \textsc{GALPY}, while the dashed lines located above correspond to the other potentials considered for our comparison. A maximum difference of $\sim 100$\,km\,s$^{-1}$ is observed between the models along the full range in $R$ included in this plot. The red points indicate the total velocities computed for the 23 HVS candidates included in our work.} 
    \label{fig:plotvesc}
\end{figure}

\subsection{Selecting a different Galactic potential}
\label{potentials}

In the previous section, we have seen that our ability to classify an object as an HVS and establish its place of origin is affected by the level of uncertainty in {\it Gaia}’s astrometry. On the other hand, we have relied exclusively on the MWPotential2014 \citep{Bovy2015}, but it is important to assess the robustness of our classification under alternative Galactic potential models  available in the literature.

For this purpose, we have used AGAMA\footnote{https://github.com/GalacticDynamics-Oxford/Agama}, which allows us to easily integrate orbits back in time with different gravitational potentials, both analytical or extracted from N-body simulations via basis function expansions \citep{Vasiliev2019}.  We considered the \cite{McMillan2017} model of the Galactic gravitational potential and also a modified version, where the potential of a bar with a total mass of $10^{10}$\,M$_{\odot}$ is added using the \cite{Pichardo2004} model with an angular velocity of $\Omega_{\rm b}$ = 35 km\,s$^{-1}$\,kpc$^{-1}$ and without modifying the total mass of the system; the MilkyWayModel contained in \textsc{GALA} \citep{Price-Whelan2017} was also included in this analysis. In order to account for the reflex motion that the existence of a massive (10$^{11}$\,M$_{\odot}$) satellite as the LMC causes in the evolution of the Galactic potential over time \citep[see discussion in ][]{Boubert2018,Vasiliev2023}, we also adopted two MW+LMC potentials: one of them which assumes fixed mass  for both systems  (LMC-frozen) and a second one where the density distributions may vary over time (LMC-live). 

We computed the past orbits for all the stars in our sample following the same procedure described above and considering the different potentials. New minimum Galactocentric distances were estimated for each of the potentials, as well as their $P_{\rm esc}$ (see Table \ref{table_results2}). Independently of the potential considered, the only star with an orbit consistent with $r \leq 1$\,kpc within uncertainties, thus being possibly originated via the Hills mechanism, is HVS07 with $r_{\rm min} \sim 1.2$\,kpc (same value obtained before). It is not surprising that these minimum distances are similar to the one previously obtained, since these fast travelers are expected to be less influenced by the MW mass distribution than other objects within the Galaxy in only a few Myr.    

However, the classification of our targets as HVSs and the resulting $P_{\rm esc}$ vary significantly depending on the choice of the  MW potential model. Only two of our targets (three when the bar is included) have $P_{\rm esc} \ge 50\%$ when the \cite{McMillan2017} model is implemented. In the Toomre diagram shown in Figure \ref{fig:plot3}, most stars lie within the contour corresponding to 600 km\,s$^{-1}$ in the ($V$, $\sqrt{U^{2}+W^{2}}$) plane. This value reflects the average $v_{\rm esc}$ in the solar neighborhood, based on that Galactic potential model. Figure \ref{fig:plotvesc} further illustrates the differences between Galaxy models, where $v_{\rm esc}$ is displayed as a function of the MW potential model in the distance range $4.5 \leq R\,[{\rm kpc}] \leq 10.0$. A mean difference of $\sim$ 100\,km\,s$^{-1}$ between the $v_{\rm esc}$ values predicted by MWPotential14 and the \cite{McMillan2017} models is found across the whole distance range. 

At an intermediate level, the prediction from the \cite{Price-Whelan2017} model is almost identical to the one obtained when considering the LMC live/frozen models, allowing 9 of our target stars to be classified as HVSs. Since the peculiar velocity of the Sun in the direction of the Galactic rotation is not particularly well constrained - variations on the order of $\sim 10$\,km\,s$^{-1}$ - its contribution is relatively minor compared to the larger uncertainties present in our total velocity estimates. Therefore, $v_{\rm T}$ values in Figure \ref{fig:plotvesc} were derived using the same Solar motion parameters for simplicity, regardless of the Galactic potential adopted.

These results show us that computing the past orbits of these HVS targets and establishing their origin is more affected by the uncertainties associated with the astrometric parameters and $v_{\rm los}$ than by the MW Galactic potential considered \citep[we also refer the reader to the recent work by ][]{Armstrong2025}. However, whether a HiVel star is truly a HVS or not critically depends on the potential, and more specifically, on the MW dark halo mass \citep[see discussion in ][]{Monari2018}.

\subsection{Prediction for the rest of HVS candidates}

\begin{table*}
\centering
\caption{Minimum Galactocentric distances and escape probabilities for the analyzed HVS candidates under different Galactic potential models. }
\begin{tabular}{lllllllllll}
\hline
 ID & $r_{\rm min,PW}$ & $P_{\rm esc,PW}$ & $r_{\rm min,MM}$ & $P_{\rm esc,MM}$ & $r_{\rm min,MM55}$ & $P_{\rm esc,MM55}$ & $r_{\rm min,LMCf}$ & $P_{\rm esc,LMCf}$ & $r_{\rm min,LMCl}$ & $P_{\rm esc,LMCl}$\\
    & (kpc) & (\%) & (kpc) & (\%) & (kpc) & (\%) & (kpc) & (\%) & (kpc) & (\%)\\
\hline
HVS01 & 9.9$^{\rm 0.3}_{\rm -0.2}$ & 99.9 & 9.9$^{\rm 0.3}_{\rm -0.2}$ & 99.6 & 9.9$^{\rm 0.3}_{\rm -0.2}$ & 99.6 & 9.9$^{\rm 0.3}_{\rm -0.2}$ & 100.0 & 9.9$^{\rm 0.2}_{\rm -0.2}$ & 100.0\\
HVS02 & 8.1$^{\rm 0.1}_{\rm -0.1}$ & 81.2 & 8.1$^{\rm 0.1}_{\rm -0.1}$ & 68.5 & 8.1$^{\rm 0.1}_{\rm -0.1}$ & 70.7 & 8.1$^{\rm 0.1}_{\rm 0.1}$ & 82.8 & 8.1$^{\rm 0.1}_{\rm -0.0}$ & 84.9\\
HVS03 & 6.5$^{\rm 0.1}_{\rm -0.2}$ & 67.4 & 6.5$^{\rm 0.1}_{\rm -0.2}$ & 49.8 & 6.5$^{\rm 0.1}_{\rm -0.2}$ & 51.1 & 6.5$^{\rm 0.2}_{\rm -0.2}$ & 76.2 & 6.5$^{\rm 0.2}_{\rm -0.2}$ & 70.6\\
HVS04 & 3.5$^{\rm 0.2}_{\rm -0.1}$ & 66.1 & 3.5$^{\rm 0.2}_{\rm -0.1}$ & 25.2 & 3.4$^{\rm 0.3}_{\rm -0.1}$ & 27.4 & 3.8$^{\rm 0.2}_{\rm -0.2}$ & 81.3 & 3.8$^{\rm 0.2}_{\rm -0.1}$ & 79.1\\
HVS05 & 7.6$^{\rm 0.1}_{\rm -0.1}$ & 61.8 & 7.5$^{\rm 0.1}_{\rm -0.1}$ & 47.0 & 7.5$^{\rm 0.1}_{\rm -0.1}$ & 49.6 & 7.6$^{\rm 0.1}_{\rm -0.1}$ & 64.9 & 7.6$^{\rm 0.1}_{\rm -0.1}$ & 67.8\\
HVS06 & 4.4$^{\rm 0.7}_{\rm -0.7}$ & 48.5 & 4.4$^{\rm 0.7}_{\rm -0.7}$ & 25.7 & 4.3$^{\rm 0.7}_{\rm -0.7}$ & 26.4 & 4.7$^{\rm 0.6}_{\rm -0.6}$ & 60.8 & 4.6$^{\rm 0.6}_{\rm -0.5}$ & 58.1\\
HVS07 & 1.3$^{\rm 0.3}_{\rm -0.2}$ & 54.1 & 1.3$^{\rm 0.3}_{\rm -0.1}$ & 39.5 & 1.3$^{\rm 0.3}_{\rm -0.2}$ & 38.4 & 1.4$^{\rm 0.3}_{\rm -0.2}$ & 67.4 & 1.4$^{\rm 0.3}_{\rm -0.2}$ & 61.6\\
HVS08 & 6.7$^{\rm 0.2}_{\rm -0.3}$ & 54.0 & 6.7$^{\rm 0.2}_{\rm -0.4}$ & 44.6 & 6.7$^{\rm 0.2}_{\rm -0.3}$ & 42.0 & 6.7$^{\rm 0.2}_{\rm -0.3}$ & 59.7 & 6.7$^{\rm 0.2}_{\rm -0.3}$ & 62.6\\
HVS09 & 3.2$^{\rm 0.3}_{\rm -0.4}$ & 46.6 & 3.2$^{\rm 0.4}_{\rm -0.4}$ & 24.9 & 3.2$^{\rm 0.3}_{\rm -0.4}$ & 23.3 & 3.4$^{\rm 0.3}_{\rm -0.3}$ & 55.8 & 3.4$^{\rm 0.3}_{\rm -0.3}$ & 57.5\\
HVS10 & 8.4$^{\rm 0.1}_{\rm 0.1}$ & 42.3 & 8.4$^{\rm 0.1}_{\rm 0.1}$ & 16.6 & 8.4$^{\rm 0.1}_{\rm 0.1}$ & 17.1 & 8.4$^{\rm 0.1}_{\rm 0.1}$ & 59.9 & 8.4$^{\rm 0.1}_{\rm 0.1}$ & 58.7\\
HVS11 & 7.3$^{\rm 0.1}_{\rm -0.1}$ & 54.0 & 7.3$^{\rm 0.1}_{\rm -0.1}$ & 31.5 & 7.3$^{\rm 0.1}_{\rm -0.1}$ & 34.4 & 7.3$^{\rm 0.1}_{\rm -0.1}$ & 62.6 & 7.3$^{\rm 0.1}_{\rm -0.1}$ & 63.1\\
HVS12 & 5.9$^{\rm 0.2}_{\rm -0.2}$ & 37.3 & 5.8$^{\rm 0.2}_{\rm -0.2}$ & 23.0 & 5.8$^{\rm 0.2}_{\rm -0.2}$ & 23.3 & 6.0$^{\rm 0.2}_{\rm -0.3}$ & 45.1 & 6.0$^{\rm 0.2}_{\rm -0.3}$ & 44.3\\
HVS13 & 5.6$^{\rm 0.1}_{\rm -0.2}$ & 41.6 & 5.6$^{\rm 0.1}_{\rm -0.2}$ & 27.4 & 5.6$^{\rm 0.1}_{\rm -0.2}$ & 27.0 & 5.8$^{\rm 0.2}_{\rm -0.3}$ & 48.6 & 5.8$^{\rm 0.2}_{\rm -0.3}$ & 48.1\\
HVS14 & 6.9$^{\rm 0.1}_{\rm 0.1}$ & 40.7 & 6.9$^{\rm 0.1}_{\rm 0.1}$ & 24.3 & 6.9$^{\rm 0.1}_{\rm 0.1}$ & 25.5 & 6.9$^{\rm 0.1}_{\rm 0.1}$ & 45.1 & 6.9$^{\rm 0.1}_{\rm 0.1}$ & 46.3\\
HVS15 & 6.3$^{\rm 0.2}_{\rm -1.0}$ & 40.8 & 6.3$^{\rm 0.2}_{\rm -0.8}$ & 27.9 & 6.3$^{\rm 0.2}_{\rm -0.8}$ & 27.4 & 6.4$^{\rm 0.2}_{\rm -0.2}$ & 46.9 & 6.4$^{\rm 0.2}_{\rm -0.2}$ & 42.6\\
HVS16 & 5.8$^{\rm 0.2}_{\rm -0.4}$ & 34.9 & 5.7$^{\rm 0.2}_{\rm -0.8}$ & 16.0 & 5.7$^{\rm 0.2}_{\rm -0.8}$ & 15.7 & 5.8$^{\rm 0.3}_{\rm -0.3}$ & 44.2 & 5.8$^{\rm 0.2}_{\rm -0.3}$ & 39.9\\
HVS17 & 7.4$^{\rm 0.1}_{\rm -5.7}$ & 37.1 & 7.4$^{\rm 0.1}_{\rm -6.2}$ & 19.5 & 7.4$^{\rm 0.1}_{\rm -6.4}$ & 20.6 & 7.4$^{\rm 0.1}_{\rm 0.1}$ & 43.3 & 7.4$^{\rm 0.1}_{\rm 0.1}$ & 39.7\\
HVS18 & 2.7$^{\rm 0.8}_{\rm -0.8}$ & 22.2 & 2.7$^{\rm 0.8}_{\rm -0.9}$ & 6.4 & 2.6$^{\rm 0.8}_{\rm -1.1}$ & 6.2 & 3.2$^{\rm 0.6}_{\rm -0.6}$ & 33.6 & 3.2$^{\rm 0.7}_{\rm -0.6}$ & 34.5\\
HVS19 & 9.1$^{\rm 0.2}_{\rm -0.1}$ & 12.0 & 9.1$^{\rm 0.2}_{\rm -0.1}$ & 1.4 & 9.1$^{\rm 0.2}_{\rm -0.1}$ & 0.9 & 9.1$^{\rm 0.2}_{\rm -0.1}$ & 23.0 & 9.1$^{\rm 0.2}_{\rm -0.1}$ & 20.9\\
HVS20 & 5.1$^{\rm 0.2}_{\rm -0.2}$ & 16.5 & 5.1$^{\rm 0.2}_{\rm -0.2}$ & 5.4 & 5.1$^{\rm 0.2}_{\rm -0.2}$ & 4.9 & 5.1$^{\rm 0.1}_{\rm -0.2}$ & 26.4 & 5.1$^{\rm 0.2}_{\rm -0.2}$ & 25.9\\
HVS21 & 7.9$^{\rm 0.1}_{\rm 0.1}$ & 5.5 & 7.9$^{\rm 0.1}_{\rm 0.1}$ & 0.1 & 7.9$^{\rm 0.1}_{\rm 0.1}$ & 0.0 & 7.9$^{\rm 0.1}_{\rm 0.1}$ & 19.0 & 7.9$^{\rm 0.1}_{\rm 0.1}$ & 17.2\\
HVS22 & 8.2$^{\rm 0.1}_{\rm 0.1}$ & 0.8 & 8.2$^{\rm 0.1}_{\rm -7.6}$ & 0.0 & 8.2$^{\rm 0.1}_{\rm -7.6}$ & 0.0 & 8.2$^{\rm 0.2}_{\rm 0.1}$ & 3.8 & 8.2$^{\rm 0.1}_{\rm 0.1}$ & 4.1\\
HVS23 & 8.3$^{\rm 0.1}_{\rm 0.1}$ & 0.5 & 8.3$^{\rm 0.1}_{\rm 0.1}$ & 0.0 & 8.3$^{\rm 0.1}_{\rm 0.1}$ & 0.0 & 8.3$^{\rm 0.1}_{\rm 0.1}$ & 6.0 & 8.3$^{\rm 0.1}_{\rm 0.1}$ & 3.3\\
\hline
\end{tabular}
\label{table_results2}
\tablefoot{Columns list: (1) Star ID, (2)-(3) minimum Galactocentric distance ($r_{\rm min,PW}$) and escape probability ($P_{\rm esc,PW}$) assuming the \cite[][PW17]{Price-Whelan2017} potential, (4)-(5) same parameters for the \cite[][MM]{McMillan2017} potential, (6)-(7) for the \cite{McMillan2017}+bar \cite{Pichardo2004} potential (MM+bar), (8)-(9) for the frozen LMC model (LMCf), and (10)-(11) for the live LMC model (LMCl). The escape probability (\%) quantifies the likelihood of a star being unbound from the MW under each potential.}
\end{table*}

\begin{figure}
    \centering
    \includegraphics[width=\columnwidth]{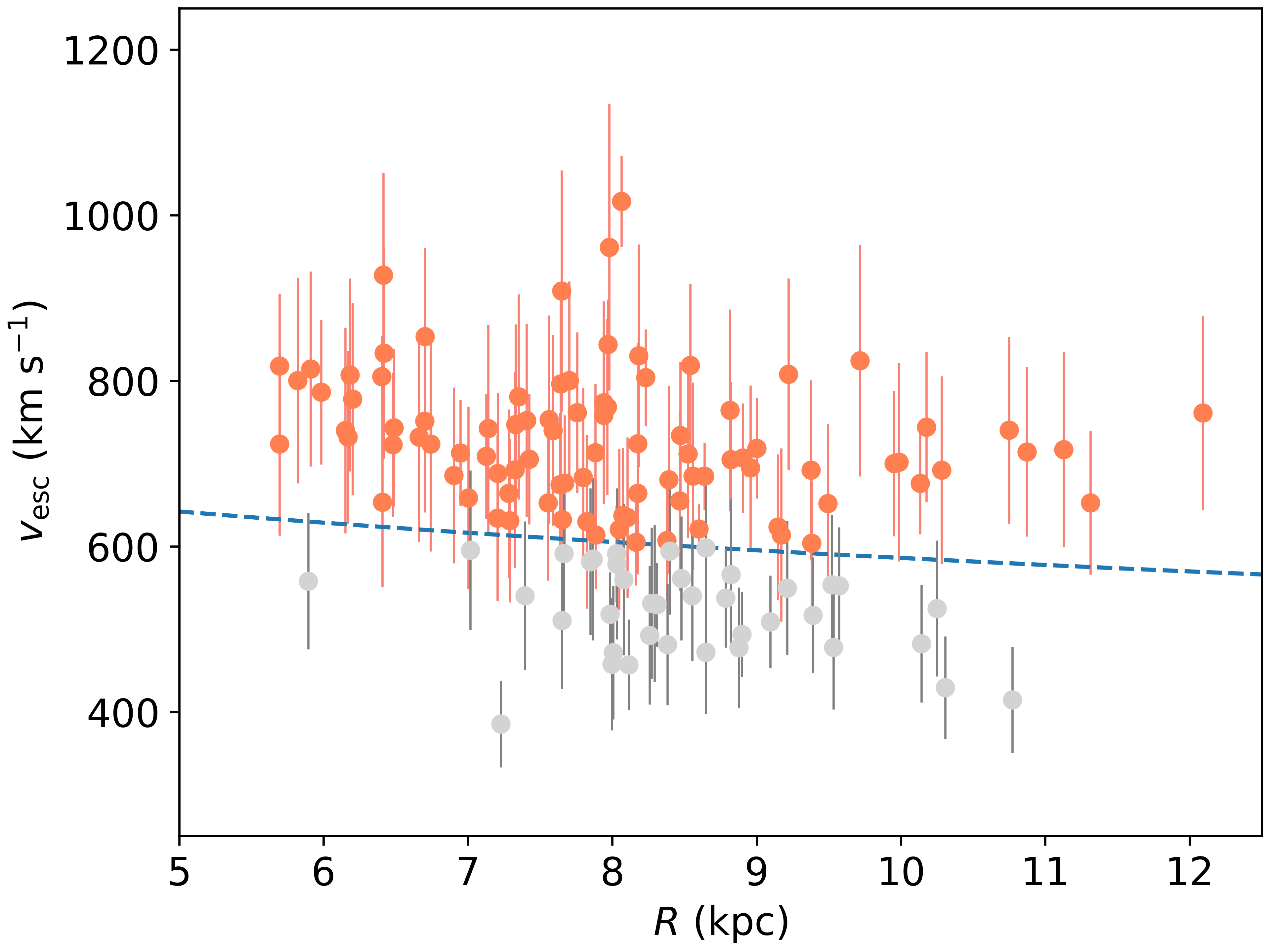}
    \caption{$v_{\rm esc}$ as a function of $R$ for the \cite{McMillan2017} MW potential (dashed blue line). The points indicate the $v_{\rm t}$ computed for the rest of candidates, where orange and gray correspond to stars with $v_{\rm t} \geq v_{\rm esc}$ and $v_{\rm t} < v_{\rm esc}$, respectively. } 
    \label{fig:plotvescv2}
\end{figure}

In this paper, we have analyzed a small fraction of the HVS candidate stars, deriving the only parameter that {\it Gaia} does not provide. This allowed us to compute  $v_{\rm T}$  and reconstruct their past orbits. However, we can extend our analysis to the remaining 132 candidate stars with $P_{\rm esc, 0} \geq 50\%$, based on their $v_{\rm t}$. Most of these stars (99) are part of the golden sample. 

As shown in Figure \ref{fig:plotvesc}, the \cite{McMillan2017} MW potential predicts a $v_{\rm esc}$ higher than the values derived from the other potentials considered in the previous section. To provide a conservative estimate of the expected number of HVSs among the remaining stars, we adopted the $v_{\rm esc}$ estimate derived from this potential as a lower limit and identified those sources that satisfy  $v_{\rm t} \geq v_{\rm esc, MM}$. Figure \ref{fig:plotvescv2} displays  $v_{\rm esc, MM}$  as a function of  R, with the $v_{\rm t}$ of the remaining candidates without spectroscopic observations yet overplotted.

Out of the sources without known $v_{\rm los}$, 93 have $v_{\rm t} \geq v_{\rm esc, MM}$, making them potential HVS candidates. Only 13 of these have previously been suggested in the literature as likely HVSs, thus the resulting number of candidates remains significant for this particular MW potential. However, the absence of measured  $v_{\rm los}$  complicates further analysis and classification based on the criteria used throughout this work. Follow-up spectroscopy will allow us to establish the nature of the remaining HVS candidates and confirm the suitability of blind searching methodologies based on DR3 or future {\it Gaia} data releases.

\section{Conclusions}

In this study, we estimate the probability for well-behaved {\it Gaia DR3} sources without known $v_{\rm los}$ of having $v_{\rm T} > v_{\rm esc}$. Among the sources with $P_{\rm esc, 0} \geq 50\% $, we analyzed a subset of 23 HVS candidates, deriving the missing $v_{\rm los}$ component from FORS2 spectra and computing their past orbits. Our results indicate that the majority of observed candidates are unlikely to originate from the Galactic center, challenging the assumption that HVSs are primarily ejected via interactions with the central SMBH through the so-called Hills mechanism. Instead, alternative mechanisms, such as satellite galaxy accretion, dynamical interactions in globular clusters, or supernova explosions in binary systems, may play a significant role in their ejection.

Although we do not completely establish the origin of these HVSs, approximately one-third of the sample  has past orbital trajectories consistent with an extra-Galactic origin. One of them, namely HVS01, is traveling through the MW at $v_{\rm T} \sim 900$\,km\,s$^{-1}$, which is remarkably high and supports the notion that its origin differs from more conventional acceleration mechanisms within the Galactic halo. Overall, their past orbits are not strongly dependent on the MW potential model considered, thus uncertainties associated with the astrometry and the $v_{\rm los}$ values derived are the main limiting factor when studying these potentially excellent tracers of the hierarchical formation of our Galaxy.

When we extended our analysis to the remaining 132 HVS candidates with an initial escape probability  $P_{\rm esc,0} > 50\%$, we identify 93 stars $v_{\rm t} \geq v_{\rm esc}$ in the Galactic potential with the largest $v_{\rm esc}$ (most  unfavorable for escaping stars). While this reinforces their potential as unbound stars, the absence of measured $v_{\rm los}$ remains a limiting factor in confirming their nature. Future spectroscopic follow-up, such as the one currently being conducted by our team, combined with improved astrometry from upcoming {\it Gaia} data releases, will be essential for refining their classification and constraining the ejection mechanisms that shape the Galactic HVS population.

\begin{acknowledgements}
We thank the anonymous referee for a careful reading of the manuscript and for constructive comments that helped improve the quality and clarity of this work.  We thank C. Mazzucchelli and M. Berton for coordinating the ``all-weather" filler programs on which this work is based. We thank C. Mateu and K. Vivas for their useful comments on the impact that the variable nature of one of our targets could have in our results. JAC-B acknowledges support from FONDECYT Regular N 1220083. PR would like to thank all the participants of GalPhases24 for the helpful comments and fruitful discussions. FG gratefully acknowledges support from the French National Research Agency (ANR) funded project ``MWDisc'' (ANR-20-CE31-0004) and ``Pristine'' (ANR-18-CE31-0017). This work has made use of data from the European Space Agency (ESA) mission
{\it Gaia} (\url{https://www.cosmos.esa.int/gaia}), processed by the {\it Gaia}
Data Processing and Analysis Consortium (DPAC,
\url{https://www.cosmos.esa.int/web/gaia/dpac/consortium}). Funding for the DPAC
has been provided by national institutions, in particular the institutions
participating in the {\it Gaia} Multilateral Agreement.
\end{acknowledgements}

\bibliographystyle{aa} 
\bibliography{biblio} 

\begin{thebibliography}{97}
\expandafter\ifx\csname natexlab\endcsname\relax\def\natexlab#1{#1}\fi

\bibitem[{{Abadi} {et~al.}(2009){Abadi}, {Navarro}, \& {Steinmetz}}]{Abadi2009}
{Abadi}, M.~G., {Navarro}, J.~F., \& {Steinmetz}, M. 2009, ApJL, 691, L63

\bibitem[{{Armstrong} {et~al.}(2025){Armstrong}, {Evans}, \&
  {Bovy}}]{Armstrong2025}
{Armstrong}, I., {Evans}, F.~A., \& {Bovy}, J. 2025, \apj, 984, 56

\bibitem[{{Bailer-Jones} {et~al.}(2021){Bailer-Jones}, {Rybizki}, {Fouesneau},
  {Demleitner}, \& {Andrae}}]{Bailer-Jones2021}
{Bailer-Jones}, C.~A.~L., {Rybizki}, J., {Fouesneau}, M., {Demleitner}, M., \&
  {Andrae}, R. 2021, AJ, 161, 147

\bibitem[{{Bamba} \& {Williams}(2022)}]{Bamba2022}
{Bamba}, A. \& {Williams}, B.~J. 2022, in Handbook of X-ray and Gamma-ray
  Astrophysics, ed. C.~{Bambi} \& A.~{Sangangelo}, 77

\bibitem[{Baumgardt {et~al.}(2006)Baumgardt, Gualandris, \&
  Zwart}]{Baumgardt2006}
Baumgardt, H., Gualandris, A., \& Zwart, S.~P. 2006, MNRAS, 372, 174

\bibitem[{{Baumgardt} \& {Vasiliev}(2021)}]{Baumgardt2021}
{Baumgardt}, H. \& {Vasiliev}, E. 2021, MNRAS, 505, 5957

\bibitem[{{Bland-Hawthorn} \& {Gerhard}(2016)}]{Bland-Hawthorn2016}
{Bland-Hawthorn}, J. \& {Gerhard}, O. 2016, ARA\&A, 54, 529

\bibitem[{Boubert {et~al.}(2018)Boubert, Guillochon, Hawkins, Ginsburg, Evans,
  \& Strader}]{Boubert2018}
Boubert, D., Guillochon, J., Hawkins, K., {et~al.} 2018, MNRAS, 479, 2789

\bibitem[{{Bovy}(2015)}]{Bovy2015}
{Bovy}, J. 2015, ApJS, 216, 29

\bibitem[{{Bovy} {et~al.}(2012){Bovy}, {Allende Prieto}, {Beers}, {Bizyaev},
  {da Costa}, {Cunha}, {Ebelke}, {Eisenstein}, {Frinchaboy}, {Garc{\'\i}a
  P{\'e}rez}, {Girardi}, {Hearty}, {Hogg}, {Holtzman}, {Maia}, {Majewski},
  {Malanushenko}, {Malanushenko}, {M{\'e}sz{\'a}ros}, {Nidever}, {O'Connell},
  {O'Donnell}, {Oravetz}, {Pan}, {Rocha-Pinto}, {Schiavon}, {Schneider},
  {Schultheis}, {Skrutskie}, {Smith}, {Weinberg}, {Wilson}, \&
  {Zasowski}}]{Bovy2012}
{Bovy}, J., {Allende Prieto}, C., {Beers}, T.~C., {et~al.} 2012, ApJ, 759, 131

\bibitem[{{Bromley} {et~al.}(2006){Bromley}, {Kenyon}, {Geller}, {Barcikowski},
  {Brown}, \& {Kurtz}}]{Bromley2006}
{Bromley}, B.~C., {Kenyon}, S.~J., {Geller}, M.~J., {et~al.} 2006, \apj, 653,
  1194

\bibitem[{{Brown}(2015)}]{Brown2015}
{Brown}, W.~R. 2015, ARA\&A, 53, 15

\bibitem[{Brown {et~al.}(2010)Brown, Anderson, Gnedin, Bond, Geller, Kenyon, \&
  Livio}]{Brown2010}
Brown, W.~R., Anderson, J., Gnedin, O.~Y., {et~al.} 2010, ApJL, 719, L23

\bibitem[{Brown {et~al.}(2014)Brown, Geller, \& Kenyon}]{Brown2014}
Brown, W.~R., Geller, M.~J., \& Kenyon, S.~J. 2014, ApJ, 787, 89

\bibitem[{Brown {et~al.}(2005)Brown, Geller, Kenyon, \& Kurtz}]{Brown2005}
Brown, W.~R., Geller, M.~J., Kenyon, S.~J., \& Kurtz, M.~J. 2005, ApJL, 622,
  L33

\bibitem[{Brown {et~al.}(2006)Brown, Geller, Kenyon, \& Kurtz}]{Brown2006}
Brown, W.~R., Geller, M.~J., Kenyon, S.~J., \& Kurtz, M.~J. 2006, ApJL, 640,
  L35

\bibitem[{{Burgasser} {et~al.}(2024){Burgasser}, {Gerasimov}, {Kremer},
  {Brooks}, {Alvarado}, {Schneider}, {Meisner}, {Theissen}, {Softich},
  {Karpoor}, {Bickle}, {Kabatnik}, {Rothermich}, {Caselden}, {Kirkpatrick},
  {Faherty}, {Casewell}, {Kuchner}, \& {The Backyard Worlds: Planet 9
  Collaboration}}]{Burgasser2024}
{Burgasser}, A.~J., {Gerasimov}, R., {Kremer}, K., {et~al.} 2024, ApJL, 971,
  L25

\bibitem[{{Cabrera} \& {Rodriguez}(2023)}]{Cabrera2023}
{Cabrera}, T. \& {Rodriguez}, C.~L. 2023, ApJ, 953, 19

\bibitem[{{Capuzzo-Dolcetta} \& {Fragione}(2015)}]{Capuzzo-Dolcetta2015}
{Capuzzo-Dolcetta}, R. \& {Fragione}, G. 2015, MNRAS, 454, 2677

\bibitem[{Chu {et~al.}(2023)Chu, Do, Ghez, Gautam, Ciurlo, O'neil, Hosek, Hees,
  Naoz, Sakai, Lu, Chen, Bentley, Becklin, \& Matthews}]{Chu2023}
Chu, D.~S., Do, T., Ghez, A., {et~al.} 2023, ApJ, 948, 94

\bibitem[{{Clementini} {et~al.}(2023){Clementini}, {Ripepi}, {Garofalo},
  {Molinaro}, {Muraveva}, {Leccia}, {Rimoldini}, {Holl}, {Jevardat de
  Fombelle}, {Sartoretti}, {Marchal}, {Audard}, {Nienartowicz}, {Andrae},
  {Marconi}, {Szabados}, {Evans}, {Lecoeur-Taibi}, {Mowlavi}, {Musella}, \&
  {Eyer}}]{Clementini2023}
{Clementini}, G., {Ripepi}, V., {Garofalo}, A., {et~al.} 2023, A\&A, 674, A18

\bibitem[{{Contigiani} {et~al.}(2019){Contigiani}, {Rossi}, \&
  {Marchetti}}]{Contigiani2019}
{Contigiani}, O., {Rossi}, E.~M., \& {Marchetti}, T. 2019, MNRAS, 487, 4025

\bibitem[{Du {et~al.}(2018)Du, Li, Newberg, Chen, Shi, Wu, \& Ma}]{Du2018}
Du, C., Li, H., Newberg, H.~J., {et~al.} 2018, ApJL, 869, L31

\bibitem[{Du {et~al.}(2019)Du, Li, Yan, Newberg, Shi, Ma, Chen, \& Wu}]{Du2019}
Du, C., Li, H., Yan, Y., {et~al.} 2019, ApJS, 244, 4

\bibitem[{{Edelmann} {et~al.}(2005){Edelmann}, {Napiwotzki}, {Heber},
  {Christlieb}, \& {Reimers}}]{Edelmann2005}
{Edelmann}, H., {Napiwotzki}, R., {Heber}, U., {Christlieb}, N., \& {Reimers},
  D. 2005, \apjl, 634, L181

\bibitem[{Erkal {et~al.}(2019)Erkal, Boubert, Gualandris, Evans, \&
  Antonini}]{Erkal2019}
Erkal, D., Boubert, D., Gualandris, A., Evans, N.~W., \& Antonini, F. 2019,
  MNRAS, 483, 2007

\bibitem[{Evans {et~al.}(2022)Evans, Marchetti, \& Rossi}]{Evans2022}
Evans, F., Marchetti, T., \& Rossi, E. 2022, MNRAS, 512, 2350

\bibitem[{{Event Horizon Telescope Collaboration} {et~al.}(2022){Event Horizon
  Telescope Collaboration}, {Akiyama}, {Alberdi}, {Alef}, {Algaba}, {Anantua},
  {Asada}, {Azulay}, {Bach}, {Baczko}, {Ball}, {Balokovi{\'c}}, {Barrett},
  {Baub{\"o}ck}, {Benson}, {Bintley}, {Blackburn}, {Blundell}, {Bouman},
  {Bower}, {Boyce}, {Bremer}, {Brinkerink}, {Brissenden}, {Britzen},
  {Broderick}, {Broguiere}, {Bronzwaer}, {Bustamante}, {Byun}, {Carlstrom},
  {Ceccobello}, {Chael}, {Chan}, {Chatterjee}, {Chatterjee}, {Chen}, {Chen},
  {Cheng}, {Cho}, {Christian}, {Conroy}, {Conway}, {Cordes}, {Crawford},
  {Crew}, {Cruz-Osorio}, {Cui}, {Davelaar}, {De Laurentis}, {Deane}, {Dempsey},
  {Desvignes}, {Dexter}, {Dhruv}, {Doeleman}, {Dougal}, {Dzib}, {Eatough},
  {Emami}, {Falcke}, {Farah}, {Fish}, {Fomalont}, {Ford}, {Fraga-Encinas},
  {Freeman}, {Friberg}, {Fromm}, {Fuentes}, {Galison}, {Gammie}, {Garc{\'\i}a},
  {Gentaz}, {Georgiev}, {Goddi}, {Gold}, {G{\'o}mez-Ruiz}, {G{\'o}mez}, {Gu},
  {Gurwell}, {Hada}, {Haggard}, {Haworth}, {Hecht}, {Hesper}, {Heumann}, {Ho},
  {Ho}, {Honma}, {Huang}, {Huang}, {Hughes}, {Ikeda}, {Impellizzeri}, {Inoue},
  {Issaoun}, {James}, {Jannuzi}, {Janssen}, {Jeter}, {Jiang},
  {Jim{\'e}nez-Rosales}, {Johnson}, {Jorstad}, {Joshi}, {Jung}, {Karami},
  {Karuppusamy}, {Kawashima}, {Keating}, {Kettenis}, {Kim}, {Kim}, {Kim},
  {Kim}, {Kino}, {Koay}, {Kocherlakota}, {Kofuji}, {Koch}, {Koyama}, {Kramer},
  {Kramer}, {Krichbaum}, {Kuo}, {La Bella}, {Lauer}, {Lee}, {Lee}, {Leung},
  {Levis}, {Li}, {Lico}, {Lindahl}, {Lindqvist}, {Lisakov}, {Liu}, {Liu},
  {Liuzzo}, {Lo}, {Lobanov}, {Loinard}, {Lonsdale}, {Lu}, {Mao}, {Marchili},
  {Markoff}, {Marrone}, {Marscher}, {Mart{\'\i}-Vidal}, {Matsushita},
  {Matthews}, {Medeiros}, {Menten}, {Michalik}, {Mizuno}, {Mizuno}, {Moran},
  {Moriyama}, {Moscibrodzka}, {M{\"u}ller}, {Mus}, {Musoke}, {Myserlis},
  {Nadolski}, {Nagai}, {Nagar}, {Nakamura}, {Narayan}, {Narayanan},
  {Natarajan}, {Nathanail}, {Fuentes}, {Neilsen}, {Neri}, {Ni}, {Noutsos},
  {Nowak}, {Oh}, {Okino}, {Olivares}, {Ortiz-Le{\'o}n}, {Oyama}, {{\"O}zel},
  {Palumbo}, {Paraschos}, {Park}, {Parsons}, {Patel}, {Pen}, {Pesce},
  {Pi{\'e}tu}, {Plambeck}, {PopStefanija}, {Porth}, {P{\"o}tzl}, {Prather},
  {Preciado-L{\'o}pez}, \& {Psaltis}}]{EHT}
{Event Horizon Telescope Collaboration}, {Akiyama}, K., {Alberdi}, A., {et~al.}
  2022, ApJL, 930, L12

\bibitem[{{Eyer} {et~al.}(2023){Eyer}, {Audard}, {Holl}, {Rimoldini},
  {Carnerero}, {Clementini}, {De Ridder}, {Distefano}, {Evans}, {Gavras},
  {Gomel}, {Lebzelter}, {Marton}, {Mowlavi}, {Panahi}, {Ripepi}, {Wyrzykowski},
  {Nienartowicz}, {Jevardat de Fombelle}, {Lecoeur-Taibi}, {Rohrbasser},
  {Riello}, {Garc{\'\i}a-Lario}, {Lanzafame}, {Mazeh}, {Raiteri}, {Zucker},
  {{\'A}brah{\'a}m}, {Aerts}, {Aguado}, {Anderson}, {Bashi}, {Binnenfeld},
  {Faigler}, {Garofalo}, {Karbevska}, {K{\'o}sp{\'a}l}, {Kruszy{\'n}ska},
  {Kun}, {Lanza}, {Leccia}, {Marconi}, {Messina}, {Molinaro}, {Moln{\'a}r},
  {Muraveva}, {Musella}, {Nagy}, {Pagano}, {Palaversa}, {Plachy}, {Pr{\v{s}}a},
  {Rybicki}, {Shahaf}, {Szabados}, {Szegedi-Elek}, {Trabucchi}, {Barblan},
  {Grenon}, {Roelens}, \& {S{\"u}veges}}]{Eyer2023}
{Eyer}, L., {Audard}, M., {Holl}, B., {et~al.} 2023, A\&A, 674, A13

\bibitem[{{Fang} {et~al.}(2025){Fang}, {Li}, \& {Li}}]{Fang2025}
{Fang}, Z., {Li}, X., \& {Li}, H. 2025, \apjs, 277, 30

\bibitem[{{Ferrand} \& {Safi-Harb}(2012)}]{Ferrand2012}
{Ferrand}, G. \& {Safi-Harb}, S. 2012, AdSpR, 49, 1313

\bibitem[{{Fiteni} {et~al.}(2021){Fiteni}, {Caruana}, {Amarante}, {Debattista},
  \& {Beraldo e Silva}}]{Fiteni2021}
{Fiteni}, K., {Caruana}, J., {Amarante}, J. A.~S., {Debattista}, V.~P., \&
  {Beraldo e Silva}, L. 2021, MNRAS, 503, 1418

\bibitem[{{Fragione} \& {Capuzzo-Dolcetta}(2016)}]{Fragione2016}
{Fragione}, G. \& {Capuzzo-Dolcetta}, R. 2016, MNRAS, 458, 2596

\bibitem[{{Fragione} \& {Gualandris}(2019)}]{Fragione2019}
{Fragione}, G. \& {Gualandris}, A. 2019, MNRAS, 489, 4543

\bibitem[{{Gaia Collaboration} {et~al.}(2018{\natexlab{a}}){Gaia
  Collaboration}, {Brown}, {Vallenari}, {Prusti}, {de Bruijne}, {Babusiaux},
  {Bailer-Jones}, {Biermann}, {Evans}, {Eyer}, {Jansen}, {Jordi}, {Klioner},
  {Lammers}, {Lindegren}, {Luri}, {Mignard}, {Panem}, {Pourbaix}, {Randich},
  {Sartoretti}, {Siddiqui}, {Soubiran}, {van Leeuwen}, {Walton}, {Arenou},
  {Bastian}, {Cropper}, {Drimmel}, {Katz}, {Lattanzi}, {Bakker}, {Cacciari},
  {Casta{\~n}eda}, {Chaoul}, {Cheek}, {De Angeli}, {Fabricius}, {Guerra},
  {Holl}, {Masana}, {Messineo}, {Mowlavi}, {Nienartowicz}, {Panuzzo},
  {Portell}, {Riello}, {Seabroke}, {Tanga}, {Th{\'e}venin}, {Gracia-Abril},
  {Comoretto}, {Garcia-Reinaldos}, {Teyssier}, {Altmann}, {Andrae}, {Audard},
  {Bellas-Velidis}, {Benson}, {Berthier}, {Blomme}, {Burgess}, {Busso},
  {Carry}, {Cellino}, {Clementini}, {Clotet}, {Creevey}, {Davidson}, {De
  Ridder}, {Delchambre}, {Dell'Oro}, {Ducourant},
  {Fern{\'a}ndez-Hern{\'a}ndez}, {Fouesneau}, {Fr{\'e}mat}, {Galluccio},
  {Garc{\'\i}a-Torres}, {Gonz{\'a}lez-N{\'u}{\~n}ez}, {Gonz{\'a}lez-Vidal},
  {Gosset}, {Guy}, {Halbwachs}, {Hambly}, {Harrison}, {Hern{\'a}ndez},
  {Hestroffer}, {Hodgkin}, {Hutton}, {Jasniewicz}, {Jean-Antoine-Piccolo},
  {Jordan}, {Korn}, {Krone-Martins}, {Lanzafame}, {Lebzelter}, {L{\"o}ffler},
  {Manteiga}, {Marrese}, {Mart{\'\i}n-Fleitas}, {Moitinho}, {Mora}, {Muinonen},
  {Osinde}, {Pancino}, {Pauwels}, {Petit}, {Recio-Blanco}, {Richards},
  {Rimoldini}, {Robin}, {Sarro}, {Siopis}, {Smith}, {Sozzetti}, {S{\"u}veges},
  {Torra}, {van Reeven}, {Abbas}, {Abreu Aramburu}, {Accart}, {Aerts},
  {Altavilla}, {{\'A}lvarez}, {Alvarez}, {Alves}, {Anderson}, {Andrei},
  {Anglada Varela}, {Antiche}, {Antoja}, {Arcay}, {Astraatmadja}, {Bach},
  {Baker}, {Balaguer-N{\'u}{\~n}ez}, {Balm}, {Barache}, {Barata}, {Barbato},
  {Barblan}, {Barklem}, {Barrado}, {Barros}, {Barstow}, {Bartholom{\'e}
  Mu{\~n}oz}, {Bassilana}, {Becciani}, {Bellazzini}, {Berihuete}, {Bertone},
  {Bianchi}, {Bienaym{\'e}}, {Blanco-Cuaresma}, {Boch}, {Boeche}, {Bombrun},
  {Borrachero}, {Bossini}, {Bouquillon}, {Bourda}, {Bragaglia}, {Bramante},
  {Breddels}, {Bressan}, {Brouillet}, {Br{\"u}semeister}, {Brugaletta},
  {Bucciarelli}, {Burlacu}, {Busonero}, {Butkevich}, {Buzzi}, {Caffau},
  {Cancelliere}, {Cannizzaro}, {Cantat-Gaudin}, {Carballo}, {Carlucci},
  {Carrasco}, {Casamiquela}, {Castellani}, {Castro-Ginard}, {Charlot},
  {Chemin}, {Chiavassa}, {Cocozza}, {Costigan}, {Cowell}, {Crifo}, {Crosta},
  {Crowley}, {Cuypers}, {Dafonte}, {Damerdji}, {Dapergolas}, {David}, {David},
  {de Laverny}, {De Luise}, {De March}, {de Martino}, {de Souza}, {de Torres},
  {Debosscher}, {del Pozo}, {Delbo}, {Delgado}, {Delgado}, {Di Matteo},
  {Diakite}, {Diener}, {Distefano}, {Dolding}, {Drazinos}, {Dur{\'a}n},
  {Edvardsson}, {Enke}, {Eriksson}, {Esquej}, {Eynard Bontemps}, {Fabre},
  {Fabrizio}, {Faigler}, {Falc{\~a}o}, {Farr{\`a}s Casas}, {Federici},
  {Fedorets}, {Fernique}, {Figueras}, {Filippi}, {Findeisen}, {Fonti},
  {Fraile}, {Fraser}, {Fr{\'e}zouls}, {Gai}, {Galleti}, {Garabato},
  {Garc{\'\i}a-Sedano}, {Garofalo}, {Garralda}, {Gavel}, {Gavras}, {Gerssen},
  {Geyer}, {Giacobbe}, {Gilmore}, {Girona}, {Giuffrida}, {Glass}, {Gomes},
  {Granvik}, {Gueguen}, {Guerrier}, {Guiraud}, {Guti{\'e}rrez-S{\'a}nchez},
  {Haigron}, {Hatzidimitriou}, {Hauser}, {Haywood}, {Heiter}, {Helmi}, {Heu},
  {Hilger}, {Hobbs}, {Hofmann}, {Holland}, {Huckle}, {Hypki}, {Icardi},
  {Jan{\ss}en}, {Jevardat de Fombelle}, {Jonker}, {Juh{\'a}sz}, {Julbe},
  {Karampelas}, {Kewley}, {Klar}, {Kochoska}, {Kohley}, {Kolenberg},
  {Kontizas}, {Kontizas}, {Koposov}, {Kordopatis}, {Kostrzewa-Rutkowska},
  {Koubsky}, {Lambert}, {Lanza}, {Lasne}, {Lavigne}, {Le Fustec}, {Le
  Poncin-Lafitte}, {Lebreton}, {Leccia}, {Leclerc}, {Lecoeur-Taibi},
  {Lenhardt}, {Leroux}, {Liao}, {Licata}, {Lindstr{\o}m}, {Lister}, {Livanou},
  {Lobel}, {L{\'o}pez}, {Managau}, {Mann}, {Mantelet}, {Marchal}, {Marchant},
  {Marconi}, {Marinoni}, {Marschalk{\'o}}, {Marshall}, {Martino}, {Marton},
  {Mary}, {Massari}, {Matijevi{\v{c}}}, {Mazeh}, {McMillan}, {Messina},
  {Michalik}, {Millar}, {Molina}, {Molinaro}, {Moln{\'a}r}, {Montegriffo},
  {Mor}, {Morbidelli}, {Morel}, {Morris}, {Mulone}, {Muraveva}, {Musella},
  {Nelemans}, {Nicastro}, {Noval}, {O'Mullane}, {Ord{\'e}novic},
  {Ord{\'o}{\~n}ez-Blanco}, {Osborne}, {Pagani}, {Pagano}, {Pailler},
  {Palacin}, {Palaversa}, {Panahi}, {Pawlak}, {Piersimoni}, {Pineau}, {Plachy},
  {Plum}, {Poggio}, {Poujoulet}, {Pr{\v{s}}a}, {Pulone}, {Racero}, {Ragaini},
  {Rambaux}, {Ramos-Lerate}, {Regibo}, {Reyl{\'e}}, {Riclet}, {Ripepi}, {Riva},
  {Rivard}, {Rixon}, {Roegiers}, {Roelens}, {Romero-G{\'o}mez}, {Rowell},
  {Royer}, {Ruiz-Dern}, {Sadowski}, {Sagrist{\`a} Sell{\'e}s}, {Sahlmann},
  {Salgado}, {Salguero}, {Sanna}, {Santana-Ros}, {Sarasso}, {Savietto},
  {Schultheis}, {Sciacca}, {Segol}, {Segovia}, {S{\'e}gransan}, {Shih},
  {Siltala}, {Silva}, {Smart}, {Smith}, {Solano}, {Solitro}, {Sordo}, {Soria
  Nieto}, {Souchay}, {Spagna}, {Spoto}, {Stampa}, {Steele},
  {Steidelm{\"u}ller}, {Stephenson}, {Stoev}, {Suess}, {Surdej}, {Szabados},
  {Szegedi-Elek}, {Tapiador}, {Taris}, {Tauran}, {Taylor}, {Teixeira},
  {Terrett}, {Teyssandier}, {Thuillot}, {Titarenko}, {Torra Clotet}, {Turon},
  {Ulla}, {Utrilla}, {Uzzi}, {Vaillant}, {Valentini}, {Valette}, {van Elteren},
  {Van Hemelryck}, {van Leeuwen}, {Vaschetto}, {Vecchiato}, {Veljanoski},
  {Viala}, {Vicente}, {Vogt}, {von Essen}, {Voss}, {Votruba}, {Voutsinas},
  {Walmsley}, {Weiler}, {Wertz}, {Wevers}, {Wyrzykowski}, {Yoldas},
  {{\v{Z}}erjal}, {Ziaeepour}, {Zorec}, {Zschocke}, {Zucker}, {Zurbach}, \&
  {Zwitter}}]{Gaia2018}
{Gaia Collaboration}, {Brown}, A.~G.~A., {Vallenari}, A., {et~al.}
  2018{\natexlab{a}}, A\&A, 616, A1

\bibitem[{{Gaia Collaboration} {et~al.}(2018{\natexlab{b}}){Gaia
  Collaboration}, {Helmi}, {van Leeuwen}, {McMillan}, {Massari}, {Antoja},
  {Robin}, {Lindegren}, {Bastian}, {Arenou}, {Babusiaux}, {Biermann},
  {Breddels}, {Hobbs}, {Jordi}, {Pancino}, {Reyl{\'e}}, {Veljanoski}, {Brown},
  {Vallenari}, {Prusti}, {de Bruijne}, {Bailer-Jones}, {Evans}, {Eyer},
  {Jansen}, {Klioner}, {Lammers}, {Luri}, {Mignard}, {Panem}, {Pourbaix},
  {Randich}, {Sartoretti}, {Siddiqui}, {Soubiran}, {Walton}, {Cropper},
  {Drimmel}, {Katz}, {Lattanzi}, {Bakker}, {Cacciari}, {Casta{\~n}eda},
  {Chaoul}, {Cheek}, {De Angeli}, {Fabricius}, {Guerra}, {Holl}, {Masana},
  {Messineo}, {Mowlavi}, {Nienartowicz}, {Panuzzo}, {Portell}, {Riello},
  {Seabroke}, {Tanga}, {Th{\'e}venin}, {Gracia-Abril}, {Comoretto},
  {Garcia-Reinaldos}, {Teyssier}, {Altmann}, {Andrae}, {Audard},
  {Bellas-Velidis}, {Benson}, {Berthier}, {Blomme}, {Burgess}, {Busso},
  {Carry}, {Cellino}, {Clementini}, {Clotet}, {Creevey}, {Davidson}, {De
  Ridder}, {Delchambre}, {Dell'Oro}, {Ducourant},
  {Fern{\'a}ndez-Hern{\'a}ndez}, {Fouesneau}, {Fr{\'e}mat}, {Galluccio},
  {Garc{\'\i}a-Torres}, {Gonz{\'a}lez-N{\'u}{\~n}ez}, {Gonz{\'a}lez-Vidal},
  {Gosset}, {Guy}, {Halbwachs}, {Hambly}, {Harrison}, {Hern{\'a}ndez},
  {Hestroffer}, {Hodgkin}, {Hutton}, {Jasniewicz}, {Jean-Antoine-Piccolo},
  {Jordan}, {Korn}, {Krone-Martins}, {Lanzafame}, {Lebzelter}, {L{\"o}ffler},
  {Manteiga}, {Marrese}, {Mart{\'\i}n-Fleitas}, {Moitinho}, {Mora}, {Muinonen},
  {Osinde}, {Pauwels}, {Petit}, {Recio-Blanco}, {Richards}, {Rimoldini},
  {Sarro}, {Siopis}, {Smith}, {Sozzetti}, {S{\"u}veges}, {Torra}, {van Reeven},
  {Abbas}, {Abreu Aramburu}, {Accart}, {Aerts}, {Altavilla}, {{\'A}lvarez},
  {Alvarez}, {Alves}, {Anderson}, {Andrei}, {Anglada Varela}, {Antiche},
  {Arcay}, {Astraatmadja}, {Bach}, {Baker}, {Balaguer-N{\'u}{\~n}ez}, {Balm},
  {Barache}, {Barata}, {Barbato}, {Barblan}, {Barklem}, {Barrado}, {Barros},
  {Barstow}, {Bartholom{\'e} Mu{\~n}oz}, {Bassilana}, {Becciani}, {Bellazzini},
  {Berihuete}, {Bertone}, {Bianchi}, {Bienaym{\'e}}, {Blanco-Cuaresma}, {Boch},
  {Boeche}, {Bombrun}, {Borrachero}, {Bossini}, {Bouquillon}, {Bourda},
  {Bragaglia}, {Bramante}, {Bressan}, {Brouillet}, {Br{\"u}semeister},
  {Brugaletta}, {Bucciarelli}, {Burlacu}, {Busonero}, {Butkevich}, {Buzzi},
  {Caffau}, {Cancelliere}, {Cannizzaro}, {Cantat-Gaudin}, {Carballo},
  {Carlucci}, {Carrasco}, {Casamiquela}, {Castellani}, {Castro-Ginard},
  {Charlot}, {Chemin}, {Chiavassa}, {Cocozza}, {Costigan}, {Cowell}, {Crifo},
  {Crosta}, {Crowley}, {Cuypers}, \& {Dafonte}}]{Gaia2018sat}
{Gaia Collaboration}, {Helmi}, A., {van Leeuwen}, F., {et~al.}
  2018{\natexlab{b}}, A\&A, 616, A12

\bibitem[{{Gaia Collaboration} {et~al.}(2016){Gaia Collaboration}, {Prusti},
  {de Bruijne}, {Brown}, {Vallenari}, {Babusiaux}, {Bailer-Jones}, {Bastian},
  {Biermann}, {Evans}, {Eyer}, {Jansen}, {Jordi}, {Klioner}, {Lammers},
  {Lindegren}, {Luri}, {Mignard}, {Milligan}, {Panem}, {Poinsignon},
  {Pourbaix}, {Randich}, {Sarri}, {Sartoretti}, {Siddiqui}, {Soubiran},
  {Valette}, {van Leeuwen}, {Walton}, {Aerts}, {Arenou}, {Cropper}, {Drimmel},
  {H{\o}g}, {Katz}, {Lattanzi}, {O'Mullane}, {Grebel}, {Holland}, {Huc},
  {Passot}, {Bramante}, {Cacciari}, {Casta{\~n}eda}, {Chaoul}, {Cheek}, {De
  Angeli}, {Fabricius}, {Guerra}, {Hern{\'a}ndez}, {Jean-Antoine-Piccolo},
  {Masana}, {Messineo}, {Mowlavi}, {Nienartowicz}, {Ord{\'o}{\~n}ez-Blanco},
  {Panuzzo}, {Portell}, {Richards}, {Riello}, {Seabroke}, {Tanga},
  {Th{\'e}venin}, {Torra}, {Els}, {Gracia-Abril}, {Comoretto},
  {Garcia-Reinaldos}, {Lock}, {Mercier}, {Altmann}, {Andrae}, {Astraatmadja},
  {Bellas-Velidis}, {Benson}, {Berthier}, {Blomme}, {Busso}, {Carry},
  {Cellino}, {Clementini}, {Cowell}, {Creevey}, {Cuypers}, {Davidson}, {De
  Ridder}, {de Torres}, {Delchambre}, {Dell'Oro}, {Ducourant}, {Fr{\'e}mat},
  {Garc{\'\i}a-Torres}, {Gosset}, {Halbwachs}, {Hambly}, {Harrison}, {Hauser},
  {Hestroffer}, {Hodgkin}, {Huckle}, {Hutton}, {Jasniewicz}, {Jordan},
  {Kontizas}, {Korn}, {Lanzafame}, {Manteiga}, {Moitinho}, {Muinonen},
  {Osinde}, {Pancino}, {Pauwels}, {Petit}, {Recio-Blanco}, {Robin}, {Sarro},
  {Siopis}, {Smith}, {Smith}, {Sozzetti}, {Thuillot}, {van Reeven}, {Viala},
  {Abbas}, {Abreu Aramburu}, {Accart}, {Aguado}, {Allan}, {Allasia},
  {Altavilla}, {{\'A}lvarez}, {Alves}, {Anderson}, {Andrei}, {Anglada Varela},
  {Antiche}, {Antoja}, {Ant{\'o}n}, {Arcay}, {Atzei}, {Ayache}, {Bach},
  {Baker}, {Balaguer-N{\'u}{\~n}ez}, {Barache}, {Barata}, {Barbier}, {Barblan},
  {Baroni}, {Barrado y Navascu{\'e}s}, {Barros}, {Barstow}, {Becciani},
  {Bellazzini}, {Bellei}, {Bello Garc{\'\i}a}, {Belokurov}, {Bendjoya},
  {Berihuete}, {Bianchi}, {Bienaym{\'e}}, {Billebaud}, {Blagorodnova},
  {Blanco-Cuaresma}, {Boch}, {Bombrun}, {Borrachero}, {Bouquillon}, {Bourda},
  {Bouy}, {Bragaglia}, {Breddels}, {Brouillet}, {Br{\"u}semeister},
  {Bucciarelli}, {Budnik}, {Burgess}, {Burgon}, {Burlacu}, {Busonero}, {Buzzi},
  {Caffau}, {Cambras}, {Campbell}, {Cancelliere}, {Cantat-Gaudin}, {Carlucci},
  {Carrasco}, {Castellani}, {Charlot}, {Charnas}, {Charvet}, {Chassat},
  {Chiavassa}, {Clotet}, {Cocozza}, {Collins}, {Collins}, {Costigan}, {Crifo},
  {Cross}, {Crosta}, {Crowley}, {Dafonte}, {Damerdji}, {Dapergolas}, {David},
  {David}, {De Cat}, {de Felice}, {de Laverny}, {De Luise}, {De March}, {de
  Martino}, {de Souza}, {Debosscher}, {del Pozo}, {Delbo}, {Delgado},
  {Delgado}, {di Marco}, {Di Matteo}, {Diakite}, {Distefano}, {Dolding}, {Dos
  Anjos}, {Drazinos}, {Dur{\'a}n}, {Dzigan}, {Ecale}, {Edvardsson}, {Enke},
  {Erdmann}, {Escolar}, {Espina}, {Evans}, {Eynard Bontemps}, {Fabre},
  {Fabrizio}, {Faigler}, {Falc{\~a}o}, {Farr{\`a}s Casas}, {Faye}, {Federici},
  {Fedorets}, {Fern{\'a}ndez-Hern{\'a}ndez}, {Fernique}, {Fienga}, {Figueras},
  {Filippi}, {Findeisen}, {Fonti}, {Fouesneau}, {Fraile}, {Fraser}, {Fuchs},
  {Furnell}, {Gai}, {Galleti}, {Galluccio}, {Garabato}, {Garc{\'\i}a-Sedano},
  {Gar{\'e}}, {Garofalo}, {Garralda}, {Gavras}, {Gerssen}, {Geyer}, {Gilmore},
  {Girona}, {Giuffrida}, {Gomes}, {Gonz{\'a}lez-Marcos},
  {Gonz{\'a}lez-N{\'u}{\~n}ez}, {Gonz{\'a}lez-Vidal}, {Granvik}, {Guerrier},
  {Guillout}, {Guiraud}, {G{\'u}rpide}, {Guti{\'e}rrez-S{\'a}nchez}, {Guy},
  {Haigron}, {Hatzidimitriou}, {Haywood}, {Heiter}, {Helmi}, {Hobbs},
  {Hofmann}, {Holl}, {Holland}, {Hunt}, {Hypki}, {Icardi}, {Irwin}, {Jevardat
  de Fombelle}, {Jofr{\'e}}, {Jonker}, {Jorissen}, {Julbe}, {Karampelas},
  {Kochoska}, {Kohley}, {Kolenberg}, {Kontizas}, {Koposov}, {Kordopatis},
  {Koubsky}, {Kowalczyk}, {Krone-Martins}, {Kudryashova}, {Kull}, {Bachchan},
  {Lacoste-Seris}, {Lanza}, {Lavigne}, {Le Poncin-Lafitte}, {Lebreton},
  {Lebzelter}, {Leccia}, {Leclerc}, {Lecoeur-Taibi}, {Lemaitre}, {Lenhardt},
  {Leroux}, {Liao}, {Licata}, {Lindstr{\o}m}, {Lister}, {Livanou}, {Lobel},
  {L{\"o}ffler}, {L{\'o}pez}, {Lopez-Lozano}, {Lorenz}, {Loureiro},
  {MacDonald}, {Magalh{\~a}es Fernandes}, {Managau}, {Mann}, {Mantelet},
  {Marchal}, {Marchant}, {Marconi}, {Marie}, {Marinoni}, {Marrese},
  {Marschalk{\'o}}, {Marshall}, {Mart{\'\i}n-Fleitas}, {Martino}, {Mary},
  {Matijevi{\v{c}}}, {Mazeh}, {McMillan}, {Messina}, {Mestre}, {Michalik},
  {Millar}, {Miranda}, {Molina}, {Molinaro}, {Molinaro}, {Moln{\'a}r},
  {Moniez}, {Montegriffo}, {Monteiro}, {Mor}, {Mora}, {Morbidelli}, {Morel},
  {Morgenthaler}, {Morley}, {Morris}, {Mulone}, {Muraveva}, {Musella},
  {Narbonne}, {Nelemans}, {Nicastro}, {Noval}, {Ord{\'e}novic},
  {Ordieres-Mer{\'e}}, {Osborne}, {Pagani}, {Pagano}, {Pailler}, {Palacin},
  {Palaversa}, {Parsons}, {Paulsen}, {Pecoraro}, {Pedrosa}, {Pentik{\"a}inen},
  {Pereira}, {Pichon}, {Piersimoni}, {Pineau}, {Plachy}, {Plum}, {Poujoulet},
  {Pr{\v{s}}a}, {Pulone}, {Ragaini}, {Rago}, {Rambaux}, {Ramos-Lerate},
  {Ranalli}, {Rauw}, {Read}, {Regibo}, {Renk}, {Reyl{\'e}}, {Ribeiro},
  {Rimoldini}, {Ripepi}, {Riva}, {Rixon}, {Roelens}, {Romero-G{\'o}mez},
  {Rowell}, {Royer}, {Rudolph}, {Ruiz-Dern}, {Sadowski}, {Sagrist{\`a}
  Sell{\'e}s}, {Sahlmann}, {Salgado}, {Salguero}, {Sarasso}, {Savietto},
  {Schnorhk}, {Schultheis}, {Sciacca}, {Segol}, {Segovia}, {Segransan},
  {Serpell}, {Shih}, {Smareglia}, {Smart}, {Smith}, {Solano}, {Solitro},
  {Sordo}, {Soria Nieto}, {Souchay}, {Spagna}, {Spoto}, {Stampa}, {Steele},
  {Steidelm{\"u}ller}, {Stephenson}, {Stoev}, {Suess}, {S{\"u}veges}, {Surdej},
  {Szabados}, {Szegedi-Elek}, {Tapiador}, {Taris}, {Tauran}, {Taylor},
  {Teixeira}, {Terrett}, {Tingley}, {Trager}, {Turon}, {Ulla}, {Utrilla},
  {Valentini}, {van Elteren}, {Van Hemelryck}, {van Leeuwen}, {Varadi},
  {Vecchiato}, {Veljanoski}, {Via}, {Vicente}, {Vogt}, {Voss}, {Votruba},
  {Voutsinas}, {Walmsley}, {Weiler}, {Weingrill}, {Werner}, {Wevers},
  {Whitehead}, {Wyrzykowski}, {Yoldas}, {{\v{Z}}erjal}, {Zucker}, {Zurbach},
  {Zwitter}, {Alecu}, {Allen}, {Allende Prieto}, {Amorim},
  {Anglada-Escud{\'e}}, {Arsenijevic}, {Azaz}, {Balm}, {Beck}, {Bernstein},
  {Bigot}, {Bijaoui}, {Blasco}, {Bonfigli}, {Bono}, {Boudreault}, {Bressan},
  {Brown}, {Brunet}, {Bunclark}, {Buonanno}, {Butkevich}, {Carret}, {Carrion},
  {Chemin}, {Ch{\'e}reau}, {Corcione}, {Darmigny}, {de Boer}, {de Teodoro}, {de
  Zeeuw}, {Delle Luche}, {Domingues}, {Dubath}, {Fodor}, {Fr{\'e}zouls},
  {Fries}, {Fustes}, {Fyfe}, {Gallardo}, {Gallegos}, {Gardiol}, {Gebran},
  {Gomboc}, {G{\'o}mez}, {Grux}, {Gueguen}, {Heyrovsky}, {Hoar}, {Iannicola},
  {Isasi Parache}, {Janotto}, {Joliet}, {Jonckheere}, {Keil}, {Kim},
  {Klagyivik}, {Klar}, {Knude}, {Kochukhov}, {Kolka}, {Kos}, {Kutka}, {Lainey},
  {LeBouquin}, {Liu}, {Loreggia}, {Makarov}, {Marseille}, {Martayan},
  {Martinez-Rubi}, {Massart}, {Meynadier}, {Mignot}, {Munari}, {Nguyen},
  {Nordlander}, {Ocvirk}, {O'Flaherty}, {Olias Sanz}, {Ortiz}, {Osorio},
  {Oszkiewicz}, {Ouzounis}, {Palmer}, {Park}, {Pasquato}, {Peltzer}, {Peralta},
  {P{\'e}turaud}, {Pieniluoma}, {Pigozzi}, {Poels}, {Prat}, {Prod'homme},
  {Raison}, {Rebordao}, {Risquez}, {Rocca-Volmerange}, {Rosen}, {Ruiz-Fuertes},
  {Russo}, {Sembay}, {Serraller Vizcaino}, {Short}, {Siebert}, {Silva},
  {Sinachopoulos}, {Slezak}, {Soffel}, {Sosnowska}, {Strai{\v{z}}ys}, {ter
  Linden}, {Terrell}, {Theil}, {Tiede}, {Troisi}, {Tsalmantza}, {Tur},
  {Vaccari}, {Vachier}, {Valles}, {Van Hamme}, {Veltz}, {Virtanen}, {Wallut},
  {Wichmann}, {Wilkinson}, {Ziaeepour}, \& {Zschocke}}]{Gaia2016}
{Gaia Collaboration}, {Prusti}, T., {de Bruijne}, J.~H.~J., {et~al.} 2016,
  A\&A, 595, A1

\bibitem[{{Gaia Collaboration} {et~al.}(2023{\natexlab{a}}){Gaia
  Collaboration}, {Recio-Blanco}, {Kordopatis}, {de Laverny}, {Palicio},
  {Spagna}, {Spina}, {Katz}, {Re Fiorentin}, {Poggio}, {McMillan}, {Vallenari},
  {Lattanzi}, {Seabroke}, {Casamiquela}, {Bragaglia}, {Antoja}, {Bailer-Jones},
  {Schultheis}, {Andrae}, {Fouesneau}, {Cropper}, {Cantat-Gaudin}, {Bijaoui},
  {Heiter}, {Brown}, {Prusti}, {de Bruijne}, {Arenou}, {Babusiaux}, {Biermann},
  {Creevey}, {Ducourant}, {Evans}, {Eyer}, {Guerra}, {Hutton}, {Jordi},
  {Klioner}, {Lammers}, {Lindegren}, {Luri}, {Mignard}, {Panem}, {Pourbaix},
  {Randich}, {Sartoretti}, {Soubiran}, {Tanga}, {Walton}, {Bastian}, {Drimmel},
  {Jansen}, {van Leeuwen}, {Bakker}, {Cacciari}, {Casta{\~n}eda}, {De Angeli},
  {Fabricius}, {Fr{\'e}mat}, {Galluccio}, {Guerrier}, {Masana}, {Messineo},
  {Mowlavi}, {Nicolas}, {Nienartowicz}, {Pailler}, {Panuzzo}, {Riclet}, {Roux},
  {Sordo}, {Th{\'e}venin}, {Gracia-Abril}, {Portell}, {Teyssier}, {Altmann},
  {Audard}, {Bellas-Velidis}, {Benson}, {Berthier}, {Blomme}, {Burgess},
  {Busonero}, {Busso}, {C{\'a}novas}, {Carry}, {Cellino}, {Cheek},
  {Clementini}, {Damerdji}, {Davidson}, {de Teodoro}, {Nu{\~n}ez Campos},
  {Delchambre}, {Dell'Oro}, {Esquej}, {Fern{\'a}ndez-Hern{\'a}ndez}, {Fraile},
  {Garabato}, {Garc{\'\i}a-Lario}, {Gosset}, {Haigron}, {Halbwachs}, {Hambly},
  {Harrison}, {Hern{\'a}ndez}, {Hestroffer}, {Hodgkin}, {Holl}, {Jan{\ss}en},
  {Jevardat de Fombelle}, {Jordan}, {Krone-Martins}, {Lanzafame},
  {L{\"o}ffler}, {Marchal}, {Marrese}, {Moitinho}, {Muinonen}, {Osborne},
  {Pancino}, {Pauwels}, {Reyl{\'e}}, {Riello}, {Rimoldini}, {Roegiers},
  {Rybizki}, {Sarro}, {Siopis}, {Smith}, {Sozzetti}, {Utrilla}, {van Leeuwen},
  {Abbas}, {{\'A}brah{\'a}m}, {Abreu Aramburu}, {Aerts}, {Aguado}, {Ajaj},
  {Aldea-Montero}, {Altavilla}, {{\'A}lvarez}, {Alves}, {Anders}, {Anderson},
  {Anglada Varela}, {Baines}, {Baker}, {Balaguer-N{\'u}{\~n}ez}, {Balbinot},
  {Balog}, {Barache}, {Barbato}, {Barros}, {Barstow}, {Bartolom{\'e}},
  {Bassilana}, {Bauchet}, {Becciani}, {Bellazzini}, {Berihuete}, {Bernet},
  {Bertone}, {Bianchi}, {Binnenfeld}, {Blanco-Cuaresma}, {Boch}, {Bombrun},
  {Bossini}, {Bouquillon}, {Bramante}, {Breedt}, {Bressan}, {Brouillet},
  {Brugaletta}, {Bucciarelli}, {Burlacu}, {Butkevich}, {Buzzi}, {Caffau},
  {Cancelliere}, {Carballo}, {Carlucci}, {Carnerero}, {Carrasco}, {Castellani},
  {Castro-Ginard}, {Chaoul}, {Charlot}, {Chemin}, {Chiaramida}, {Chiavassa},
  {Chornay}, {Comoretto}, {Contursi}, {Cooper}, {Cornez}, {Cowell}, {Crifo},
  {Crosta}, {Crowley}, {Dafonte}, {Dapergolas}, {David}, {De Luise}, {De
  March}, {De Ridder}, {de Souza}, {de Torres}, {del Peloso}, {del Pozo},
  {Delbo}, {Delgado}, {Delisle}, {Demouchy}, {Dharmawardena}, {Di Matteo},
  {Diakite}, {Diener}, {Distefano}, {Dolding}, {Edvardsson}, {Enke}, {Fabre},
  {Fabrizio}, {Faigler}, {Fedorets}, {Fernique}, {Figueras}, {Fournier},
  {Fouron}, {Fragkoudi}, {Gai}, {Garcia-Gutierrez}, {Garcia-Reinaldos},
  {Garc{\'\i}a-Torres}, {Garofalo}, {Gavel}, {Gavras}, {Gerlach}, {Geyer},
  {Giacobbe}, {Gilmore}, {Girona}, {Giuffrida}, {Gomel}, {Gomez},
  {Gonz{\'a}lez-N{\'u}{\~n}ez}, {Gonz{\'a}lez-Santamar{\'\i}a},
  {Gonz{\'a}lez-Vidal}, {Granvik}, {Guillout}, {Guiraud},
  {Guti{\'e}rrez-S{\'a}nchez}, {Guy}, {Hatzidimitriou}, {Hauser}, {Haywood},
  {Helmer}, {Helmi}, {Sarmiento}, {Hidalgo}, {H{\l}adczuk}, {Hobbs}, {Holland},
  {Huckle}, {Jardine}, {Jasniewicz}, {Jean-Antoine Piccolo},
  {Jim{\'e}nez-Arranz}, {Juaristi Campillo}, {Julbe}, {Karbevska}, {Kervella},
  {Khanna}, {Korn}, {K{\'o}sp{\'a}l}, {Kostrzewa-Rutkowska}, {Kruszy{\'n}ska},
  {Kun}, {Laizeau}, {Lambert}, {Lanza}, {Lasne}, {Le Campion}, {Lebreton},
  {Lebzelter}, {Leccia}, {Leclerc}, {Lecoeur-Taibi}, {Liao}, {Licata},
  {Lindstr{\o}m}, {Lister}, {Livanou}, {Lobel}, {Lorca}, {Loup}, {Madrero
  Pardo}, {Magdaleno Romeo}, {Managau}, {Mann}, {Manteiga}, {Marchant},
  {Marconi}, {Marcos}, {Marcos Santos}, {Mar{\'\i}n Pina}, {Marinoni},
  {Marocco}, {Marshall}, {Martin Polo}, {Mart{\'\i}n-Fleitas}, {Marton},
  {Mary}, {Masip}, {Massari}, {Mastrobuono-Battisti}, {Mazeh}, {Messina},
  {Michalik}, {Millar}, {Mints}, {Molina}, {Molinaro}, {Moln{\'a}r}, {Monari},
  {Mongui{\'o}}, {Montegriffo}, {Montero}, {Mor}, {Mora}, {Morbidelli},
  {Morel}, {Morris}, {Muraveva}, {Murphy}, {Musella}, {Nagy}, {Noval},
  {Oca{\~n}a}, {Ogden}, {Ordenovic}, {Osinde}, {Pagani}, {Pagano}, {Palaversa},
  {Pallas-Quintela}, {Panahi}, {Payne-Wardenaar}, {Pe{\~n}alosa Esteller},
  {Penttil{\"a}}, {Pichon}, {Piersimoni}, {Pineau}, {Plachy}, {Plum},
  {Pr{\v{s}}a}, {Pulone}, {Racero}, {Ragaini}, {Rainer}, {Raiteri}, {Ramos},
  {Ramos-Lerate}, {Regibo}, {Richards}, {Rios Diaz}, {Ripepi}, {Riva}, {Rix},
  {Rixon}, {Robichon}, {Robin}, {Robin}, {Roelens}, {Rogues}, {Rohrbasser},
  {Romero-G{\'o}mez}, {Rowell}, {Royer}, {Ruz Mieres}, {Rybicki}, {Sadowski},
  {S{\'a}ez N{\'u}{\~n}ez}, {Sagrist{\`a} Sell{\'e}s}, {Sahlmann}, {Salguero},
  {Samaras}, {Sanchez Gimenez}, {Sanna}, {Santove{\~n}a}, {Sarasso}, {Sciacca},
  {Segol}, {Segovia}, {S{\'e}gransan}, {Semeux}, {Shahaf}, {Siddiqui},
  {Siebert}, {Siltala}, {Silvelo}, {Slezak}, {Slezak}, {Smart}, {Snaith},
  {Solano}, {Solitro}, {Souami}, {Souchay}, {Spoto}, {Steele},
  {Steidelm{\"u}ller}, {Stephenson}, {S{\"u}veges}, {Surdej}, {Szabados},
  {Szegedi-Elek}, {Taris}, {Taylor}, {Teixeira}, {Tolomei}, {Tonello}, {Torra},
  {Torra}, {Torralba Elipe}, {Trabucchi}, {Tsounis}, {Turon}, {Ulla}, {Unger},
  {Vaillant}, {van Dillen}, {van Reeven}, {Vanel}, {Vecchiato}, {Viala},
  {Vicente}, {Voutsinas}, {Weiler}, {Wevers}, {Wyrzykowski}, {Yoldas}, {Yvard},
  {Zhao}, {Zorec}, {Zucker}, \& {Zwitter}}]{Gaia_cartography}
{Gaia Collaboration}, {Recio-Blanco}, A., {Kordopatis}, G., {et~al.}
  2023{\natexlab{a}}, \aap, 674, A38

\bibitem[{{Gaia Collaboration} {et~al.}(2023{\natexlab{b}}){Gaia
  Collaboration}, {Vallenari}, {Brown}, {Prusti}, {de Bruijne}, {Arenou},
  {Babusiaux}, {Biermann}, {Creevey}, {Ducourant}, {Evans}, {Eyer}, {Guerra},
  {Hutton}, {Jordi}, {Klioner}, {Lammers}, {Lindegren}, {Luri}, {Mignard},
  {Panem}, {Pourbaix}, {Randich}, {Sartoretti}, {Soubiran}, {Tanga}, {Walton},
  {Bailer-Jones}, {Bastian}, {Drimmel}, {Jansen}, {Katz}, {Lattanzi}, {van
  Leeuwen}, {Bakker}, {Cacciari}, {Casta{\~n}eda}, {De Angeli}, {Fabricius},
  {Fouesneau}, {Fr{\'e}mat}, {Galluccio}, {Guerrier}, {Heiter}, {Masana},
  {Messineo}, {Mowlavi}, {Nicolas}, {Nienartowicz}, {Pailler}, {Panuzzo},
  {Riclet}, {Roux}, {Seabroke}, {Sordo}, {Th{\'e}venin}, {Gracia-Abril},
  {Portell}, {Teyssier}, {Altmann}, {Andrae}, {Audard}, {Bellas-Velidis},
  {Benson}, {Berthier}, {Blomme}, {Burgess}, {Busonero}, {Busso},
  {C{\'a}novas}, {Carry}, {Cellino}, {Cheek}, {Clementini}, {Damerdji},
  {Davidson}, {de Teodoro}, {Nu{\~n}ez Campos}, {Delchambre}, {Dell'Oro},
  {Esquej}, {Fern{\'a}ndez-Hern{\'a}ndez}, {Fraile}, {Garabato},
  {Garc{\'\i}a-Lario}, {Gosset}, {Haigron}, {Halbwachs}, {Hambly}, {Harrison},
  {Hern{\'a}ndez}, {Hestroffer}, {Hodgkin}, {Holl}, {Jan{\ss}en}, {Jevardat de
  Fombelle}, {Jordan}, {Krone-Martins}, {Lanzafame}, {L{\"o}ffler}, {Marchal},
  {Marrese}, {Moitinho}, {Muinonen}, {Osborne}, {Pancino}, {Pauwels},
  {Recio-Blanco}, {Reyl{\'e}}, {Riello}, {Rimoldini}, {Roegiers}, {Rybizki},
  {Sarro}, {Siopis}, {Smith}, {Sozzetti}, {Utrilla}, {van Leeuwen}, {Abbas},
  {{\'A}brah{\'a}m}, {Abreu Aramburu}, {Aerts}, {Aguado}, {Ajaj},
  {Aldea-Montero}, {Altavilla}, {{\'A}lvarez}, {Alves}, {Anders}, {Anderson},
  {Anglada Varela}, {Antoja}, {Baines}, {Baker}, {Balaguer-N{\'u}{\~n}ez},
  {Balbinot}, {Balog}, {Barache}, {Barbato}, {Barros}, {Barstow},
  {Bartolom{\'e}}, {Bassilana}, {Bauchet}, {Becciani}, {Bellazzini},
  {Berihuete}, {Bernet}, {Bertone}, {Bianchi}, {Binnenfeld}, {Blanco-Cuaresma},
  {Blazere}, {Boch}, {Bombrun}, {Bossini}, {Bouquillon}, {Bragaglia},
  {Bramante}, {Breedt}, {Bressan}, {Brouillet}, {Brugaletta}, {Bucciarelli},
  {Burlacu}, {Butkevich}, {Buzzi}, {Caffau}, {Cancelliere}, {Cantat-Gaudin},
  {Carballo}, {Carlucci}, {Carnerero}, {Carrasco}, {Casamiquela}, {Castellani},
  {Castro-Ginard}, {Chaoul}, {Charlot}, {Chemin}, {Chiaramida}, {Chiavassa},
  {Chornay}, {Comoretto}, {Contursi}, {Cooper}, {Cornez}, {Cowell}, {Crifo},
  {Cropper}, {Crosta}, {Crowley}, {Dafonte}, {Dapergolas}, {David}, {David},
  {de Laverny}, {De Luise}, {De March}, {De Ridder}, {de Souza}, {de Torres},
  {del Peloso}, {del Pozo}, {Delbo}, {Delgado}, {Delisle}, {Demouchy},
  {Dharmawardena}, {Di Matteo}, {Diakite}, {Diener}, {Distefano}, {Dolding},
  {Edvardsson}, {Enke}, {Fabre}, {Fabrizio}, {Faigler}, {Fedorets}, {Fernique},
  {Fienga}, {Figueras}, {Fournier}, {Fouron}, {Fragkoudi}, {Gai},
  {Garcia-Gutierrez}, {Garcia-Reinaldos}, {Garc{\'\i}a-Torres}, {Garofalo},
  {Gavel}, {Gavras}, {Gerlach}, {Geyer}, {Giacobbe}, {Gilmore}, {Girona},
  {Giuffrida}, {Gomel}, {Gomez}, {Gonz{\'a}lez-N{\'u}{\~n}ez},
  {Gonz{\'a}lez-Santamar{\'\i}a}, {Gonz{\'a}lez-Vidal}, {Granvik}, {Guillout},
  {Guiraud}, {Guti{\'e}rrez-S{\'a}nchez}, {Guy}, {Hatzidimitriou}, {Hauser},
  {Haywood}, {Helmer}, {Helmi}, {Sarmiento}, {Hidalgo}, {Hilger},
  {H{\l}adczuk}, {Hobbs}, {Holland}, {Huckle}, {Jardine}, {Jasniewicz},
  {Jean-Antoine Piccolo}, {Jim{\'e}nez-Arranz}, {Jorissen}, {Juaristi
  Campillo}, {Julbe}, {Karbevska}, {Kervella}, {Khanna}, {Kontizas},
  {Kordopatis}, {Korn}, {K{\'o}sp{\'a}l}, {Kostrzewa-Rutkowska},
  {Kruszy{\'n}ska}, {Kun}, {Laizeau}, {Lambert}, {Lanza}, {Lasne}, {Le
  Campion}, {Lebreton}, {Lebzelter}, {Leccia}, {Leclerc}, {Lecoeur-Taibi},
  {Liao}, {Licata}, {Lindstr{\o}m}, {Lister}, {Livanou}, {Lobel}, {Lorca},
  {Loup}, {Madrero Pardo}, {Magdaleno Romeo}, {Managau}, {Mann}, {Manteiga},
  {Marchant}, {Marconi}, {Marcos}, {Marcos Santos}, {Mar{\'\i}n Pina},
  {Marinoni}, {Marocco}, {Marshall}, {Martin Polo}, {Mart{\'\i}n-Fleitas},
  {Marton}, {Mary}, {Masip}, {Massari}, {Mastrobuono-Battisti}, {Mazeh},
  {McMillan}, {Messina}, {Michalik}, {Millar}, {Mints}, {Molina}, {Molinaro},
  {Moln{\'a}r}, {Monari}, {Mongui{\'o}}, {Montegriffo}, {Montero}, {Mor},
  {Mora}, {Morbidelli}, {Morel}, {Morris}, {Muraveva}, {Murphy}, {Musella},
  {Nagy}, {Noval}, {Oca{\~n}a}, {Ogden}, {Ordenovic}, {Osinde}, {Pagani},
  {Pagano}, {Palaversa}, {Palicio}, {Pallas-Quintela}, {Panahi},
  {Payne-Wardenaar}, {Pe{\~n}alosa Esteller}, {Penttil{\"a}}, {Pichon},
  {Piersimoni}, {Pineau}, {Plachy}, {Plum}, {Poggio}, {Pr{\v{s}}a}, {Pulone},
  {Racero}, {Ragaini}, {Rainer}, {Raiteri}, {Rambaux}, {Ramos}, {Ramos-Lerate},
  {Re Fiorentin}, {Regibo}, {Richards}, {Rios Diaz}, {Ripepi}, {Riva}, {Rix},
  {Rixon}, {Robichon}, {Robin}, {Robin}, {Roelens}, {Rogues}, {Rohrbasser},
  {Romero-G{\'o}mez}, {Rowell}, {Royer}, {Ruz Mieres}, {Rybicki}, {Sadowski},
  {S{\'a}ez N{\'u}{\~n}ez}, {Sagrist{\`a} Sell{\'e}s}, {Sahlmann}, {Salguero},
  {Samaras}, {Sanchez Gimenez}, {Sanna}, {Santove{\~n}a}, {Sarasso},
  {Schultheis}, {Sciacca}, {Segol}, {Segovia}, {S{\'e}gransan}, {Semeux},
  {Shahaf}, {Siddiqui}, {Siebert}, {Siltala}, {Silvelo}, {Slezak}, {Slezak},
  {Smart}, {Snaith}, {Solano}, {Solitro}, {Souami}, {Souchay}, {Spagna},
  {Spina}, {Spoto}, {Steele}, {Steidelm{\"u}ller}, {Stephenson}, {S{\"u}veges},
  {Surdej}, {Szabados}, {Szegedi-Elek}, {Taris}, {Taylor}, {Teixeira},
  {Tolomei}, {Tonello}, {Torra}, {Torra}, {Torralba Elipe}, {Trabucchi},
  {Tsounis}, {Turon}, {Ulla}, {Unger}, {Vaillant}, {van Dillen}, {van Reeven},
  {Vanel}, {Vecchiato}, {Viala}, {Vicente}, {Voutsinas}, {Weiler}, {Wevers},
  {Wyrzykowski}, {Yoldas}, {Yvard}, {Zhao}, {Zorec}, {Zucker}, \&
  {Zwitter}}]{Gaia2023}
{Gaia Collaboration}, {Vallenari}, A., {Brown}, A.~G.~A., {et~al.}
  2023{\natexlab{b}}, A\&A, 674, A1

\bibitem[{Gallo {et~al.}(2022)Gallo, Ostorero, Chakrabarty, Ebagezio, \&
  Diaferio}]{Gallo2022}
Gallo, A., Ostorero, L., Chakrabarty, S.~S., Ebagezio, S., \& Diaferio, A.
  2022, A\&A, 663, A72

\bibitem[{Generozov(2020)}]{Generozov2020}
Generozov, A. 2020, ApJ, 904, 118

\bibitem[{{Giesers} {et~al.}(2018){Giesers}, {Dreizler}, {Husser}, {Kamann},
  {Anglada Escud{\'e}}, {Brinchmann}, {Carollo}, {Roth}, {Weilbacher}, \&
  {Wisotzki}}]{Giesers2018}
{Giesers}, B., {Dreizler}, S., {Husser}, T.-O., {et~al.} 2018, MNRAS, 475, L15

\bibitem[{Gualandris \& Zwart(2007)}]{Gualandris2007}
Gualandris, A. \& Zwart, S.~P. 2007, MNRAS, 376, L29

\bibitem[{{G{\"u}lzow} {et~al.}(2024){G{\"u}lzow}, {Fairbairn}, \&
  {Schwarz}}]{Gulzow2023}
{G{\"u}lzow}, L., {Fairbairn}, M., \& {Schwarz}, D.~J. 2024, MNRAS, 529, 3816

\bibitem[{{H{\"a}berle} {et~al.}(2024){H{\"a}berle}, {Neumayer}, {Seth},
  {Bellini}, {Libralato}, {Baumgardt}, {Whitaker}, {Dumont}, {Alfaro-Cuello},
  {Anderson}, {Clontz}, {Kacharov}, {Kamann}, {Feldmeier-Krause}, {Milone},
  {Nitschai}, {Pechetti}, \& {van de Ven}}]{Haberle2024}
{H{\"a}berle}, M., {Neumayer}, N., {Seth}, A., {et~al.} 2024, Nat, 631, 285

\bibitem[{{Han} {et~al.}(2025){Han}, {El-Badry}, {Lucchini}, {Hernquist},
  {Brown}, {Garavito-Camargo}, {Conroy}, \& {Sari}}]{Han2025}
{Han}, J.~J., {El-Badry}, K., {Lucchini}, S., {et~al.} 2025, \apj, 982, 188

\bibitem[{{Hattori} {et~al.}(2025){Hattori}, {Taniguchi}, {Tsujimoto},
  {Matsunaga}, {Sameshima}, {Elgueta}, \& {Otsubo}}]{Hattori2025}
{Hattori}, K., {Taniguchi}, D., {Tsujimoto}, T., {et~al.} 2025, arXiv e-prints,
  arXiv:2502.20266

\bibitem[{{Hattori} {et~al.}(2018){Hattori}, {Valluri}, {Bell}, \&
  {Roederer}}]{Hattori2018}
{Hattori}, K., {Valluri}, M., {Bell}, E.~F., \& {Roederer}, I.~U. 2018, \apj,
  866, 121

\bibitem[{{Helmi}(2020)}]{Helmi2020}
{Helmi}, A. 2020, \araa, 58, 205

\bibitem[{{Hills}(1988)}]{Hills1988}
{Hills}, J.~G. 1988, Nat, 331, 687

\bibitem[{Hirsch {et~al.}(2005)Hirsch, Heber, O'Toole, \&
  Bresolin}]{Hirsch2005}
Hirsch, H., Heber, U., O'Toole, S., \& Bresolin, F. 2005, A\&A, 444, L61

\bibitem[{Huang {et~al.}(2021)Huang, Li, Zhang, Li, Sun, Chang, Dong, \&
  Liu}]{Huang2021}
Huang, Y., Li, Q., Zhang, H., {et~al.} 2021, ApJL, 907, L42

\bibitem[{{Irrgang} {et~al.}(2021){Irrgang}, {Dimpel}, {Heber}, \&
  {Raddi}}]{Irrgang2021}
{Irrgang}, A., {Dimpel}, M., {Heber}, U., \& {Raddi}, R. 2021, \aap, 646, L4

\bibitem[{{Irrgang} {et~al.}(2018){Irrgang}, {Kreuzer}, \&
  {Heber}}]{Irrgang2018}
{Irrgang}, A., {Kreuzer}, S., \& {Heber}, U. 2018, \aap, 620, A48

\bibitem[{Koposov {et~al.}(2020)Koposov, Boubert, Li, Erkal, Costa, Zucker, Ji,
  Kuehn, Lewis, Mackey, Simpson, Shipp, Wan, Belokurov, Bland-Hawthorn,
  Martell, Nordlander, Pace, Silva, Wang, \& Collaboration}]{Koposov2020}
Koposov, S.~E., Boubert, D., Li, T.~S., {et~al.} 2020, MNRAS, 491, 2465

\bibitem[{{Koppelman} {et~al.}(2019){Koppelman}, {Helmi}, {Massari},
  {Price-Whelan}, \& {Starkenburg}}]{Koppelman2019}
{Koppelman}, H.~H., {Helmi}, A., {Massari}, D., {Price-Whelan}, A.~M., \&
  {Starkenburg}, T.~K. 2019, A\&A, 631, L9

\bibitem[{{Kreuzer} {et~al.}(2020){Kreuzer}, {Irrgang}, \&
  {Heber}}]{Kreuzer2020}
{Kreuzer}, S., {Irrgang}, A., \& {Heber}, U. 2020, \aap, 637, A53

\bibitem[{Li {et~al.}(2022)Li, Du, Ma, Shi, Newberg, \& Piao}]{Li2022}
Li, H., Du, C., Ma, J., {et~al.} 2022, ApJL, 933, L13

\bibitem[{{Li} {et~al.}(2018){Li}, {Tan}, \& {Zhao}}]{Li2018}
{Li}, H., {Tan}, K., \& {Zhao}, G. 2018, \apjs, 238, 16

\bibitem[{Li {et~al.}(2021)Li, Luo, Lu, Zhang, Li, Wang, Zuo, Xiang, Ting,
  Marchetti, Li, Wang, Zhang, Hattori, Zhao, Zhang, \& Zhao}]{Li2021}
Li, Y.-B., Luo, A.~L., Lu, Y.-J., {et~al.} 2021, ApJS, 252, 3

\bibitem[{{Lian} {et~al.}(2024){Lian}, {Zasowski}, {Chen}, {Imig}, {Wang},
  {Boardman}, \& {Liu}}]{Lian2024}
{Lian}, J., {Zasowski}, G., {Chen}, B., {et~al.} 2024, NatAs, 8, 1302

\bibitem[{Liao {et~al.}(2023)Liao, Du, Li, Ma, \& Shi}]{Liao2023}
Liao, J., Du, C., Li, H., Ma, J., \& Shi, J. 2023, ApJL, 944, L39

\bibitem[{Lin {et~al.}(2023)Lin, Xu, Hao, Li, Liu, \& Bian}]{Lin2023}
Lin, Z., Xu, Y., Hao, C., {et~al.} 2023, ApJ, 952, 64

\bibitem[{{Lindegren} {et~al.}(2021{\natexlab{a}}){Lindegren}, {Bastian},
  {Biermann}, {Bombrun}, {de Torres}, {Gerlach}, {Geyer}, {Hern{\'a}ndez},
  {Hilger}, {Hobbs}, {Klioner}, {Lammers}, {McMillan}, {Ramos-Lerate},
  {Steidelm{\"u}ller}, {Stephenson}, \& {van Leeuwen}}]{Lindegren2021parallax}
{Lindegren}, L., {Bastian}, U., {Biermann}, M., {et~al.} 2021{\natexlab{a}},
  A\&A, 649, A4

\bibitem[{{Lindegren} {et~al.}(2021{\natexlab{b}}){Lindegren}, {Klioner},
  {Hern{\'a}ndez}, {Bombrun}, {Ramos-Lerate}, {Steidelm{\"u}ller}, {Bastian},
  {Biermann}, {de Torres}, {Gerlach}, {Geyer}, {Hilger}, {Hobbs}, {Lammers},
  {McMillan}, {Stephenson}, {Casta{\~n}eda}, {Davidson}, {Fabricius},
  {Gracia-Abril}, {Portell}, {Rowell}, {Teyssier}, {Torra}, {Bartolom{\'e}},
  {Clotet}, {Garralda}, {Gonz{\'a}lez-Vidal}, {Torra}, {Abbas}, {Altmann},
  {Anglada Varela}, {Balaguer-N{\'u}{\~n}ez}, {Balog}, {Barache}, {Becciani},
  {Bernet}, {Bertone}, {Bianchi}, {Bouquillon}, {Brown}, {Bucciarelli},
  {Busonero}, {Butkevich}, {Buzzi}, {Cancelliere}, {Carlucci}, {Charlot},
  {Cioni}, {Crosta}, {Crowley}, {del Peloso}, {del Pozo}, {Drimmel}, {Esquej},
  {Fienga}, {Fraile}, {Gai}, {Garcia-Reinaldos}, {Guerra}, {Hambly}, {Hauser},
  {Jan{\ss}en}, {Jordan}, {Kostrzewa-Rutkowska}, {Lattanzi}, {Liao}, {Licata},
  {Lister}, {L{\"o}ffler}, {Marchant}, {Masip}, {Mignard}, {Mints}, {Molina},
  {Mora}, {Morbidelli}, {Murphy}, {Pagani}, {Panuzzo}, {Pe{\~n}alosa Esteller},
  {Poggio}, {Re Fiorentin}, {Riva}, {Sagrist{\`a} Sell{\'e}s}, {Sanchez
  Gimenez}, {Sarasso}, {Sciacca}, {Siddiqui}, {Smart}, {Souami}, {Spagna},
  {Steele}, {Taris}, {Utrilla}, {van Reeven}, \& {Vecchiato}}]{Lindegren2021}
{Lindegren}, L., {Klioner}, S.~A., {Hern{\'a}ndez}, J., {et~al.}
  2021{\natexlab{b}}, A\&A, 649, A2

\bibitem[{{Mackereth} {et~al.}(2019){Mackereth}, {Schiavon}, {Pfeffer},
  {Hayes}, {Bovy}, {Anguiano}, {Allende Prieto}, {Hasselquist}, {Holtzman},
  {Johnson}, {Majewski}, {O'Connell}, {Shetrone}, {Tissera}, \&
  {Fern{\'a}ndez-Trincado}}]{Mackereth2019}
{Mackereth}, J.~T., {Schiavon}, R.~P., {Pfeffer}, J., {et~al.} 2019, \mnras,
  482, 3426

\bibitem[{Marchetti {et~al.}(2018)Marchetti, Contigiani, Rossi, Albert, Brown,
  \& Sesana}]{Marchetti2018}
Marchetti, T., Contigiani, O., Rossi, E., {et~al.} 2018, MNRAS, 476, 4697

\bibitem[{Marchetti {et~al.}(2022)Marchetti, Evans, \& Rossi}]{Marchetti2022}
Marchetti, T., Evans, F.~A., \& Rossi, E.~M. 2022, MNRAS, 515, 767

\bibitem[{Marchetti {et~al.}(2019)Marchetti, Rossi, \& Brown}]{Marchetti2019}
Marchetti, T., Rossi, E., \& Brown, A. 2019, MNRAS, 490, 157

\bibitem[{Marchetti {et~al.}(2017)Marchetti, Rossi, Kordopatis, Brown, Rimoldi,
  Starkenburg, Youakim, \& Ashley}]{Marchetti2017}
Marchetti, T., Rossi, E., Kordopatis, G., {et~al.} 2017, MNRAS, 470, 1388

\bibitem[{{Matsuno} {et~al.}(2019){Matsuno}, {Aoki}, \& {Suda}}]{Matsuno2019}
{Matsuno}, T., {Aoki}, W., \& {Suda}, T. 2019, ApJL, 874, L35

\bibitem[{{McConnachie}(2012)}]{McConnachie2012}
{McConnachie}, A.~W. 2012, AJ, 144, 4

\bibitem[{{McMillan}(2017)}]{McMillan2017}
{McMillan}, P.~J. 2017, MNRAS, 465, 76

\bibitem[{{Monari} {et~al.}(2018){Monari}, {Famaey}, {Carrillo}, {Piffl},
  {Steinmetz}, {Wyse}, {Anders}, {Chiappini}, \& {Jan{\ss}en}}]{Monari2018}
{Monari}, G., {Famaey}, B., {Carrillo}, I., {et~al.} 2018, \aap, 616, L9

\bibitem[{{Montanari} {et~al.}(2019){Montanari}, {Barrado}, \&
  {Garc{\'\i}a-Bellido}}]{Montanari2019}
{Montanari}, F., {Barrado}, D., \& {Garc{\'\i}a-Bellido}, J. 2019, MNRAS, 490,
  5647

\bibitem[{{Myeong} {et~al.}(2019){Myeong}, {Vasiliev}, {Iorio}, {Evans}, \&
  {Belokurov}}]{Myeong2019}
{Myeong}, G.~C., {Vasiliev}, E., {Iorio}, G., {Evans}, N.~W., \& {Belokurov},
  V. 2019, MNRAS, 488, 1235

\bibitem[{{Palladino} {et~al.}(2014){Palladino}, {Schlesinger},
  {Holley-Bockelmann}, {Allende Prieto}, {Beers}, {Lee}, \&
  {Schneider}}]{Palladino2014}
{Palladino}, L.~E., {Schlesinger}, K.~J., {Holley-Bockelmann}, K., {et~al.}
  2014, \apj, 780, 7

\bibitem[{{Pichardo} {et~al.}(2004){Pichardo}, {Martos}, \&
  {Moreno}}]{Pichardo2004}
{Pichardo}, B., {Martos}, M., \& {Moreno}, E. 2004, ApJ, 609, 144

\bibitem[{{Pietrzy{\'n}ski} {et~al.}(2019){Pietrzy{\'n}ski}, {Graczyk},
  {Gallenne}, {Gieren}, {Thompson}, {Pilecki}, {Karczmarek}, {G{\'o}rski},
  {Suchomska}, {Taormina}, {Zgirski}, {Wielg{\'o}rski}, {Ko{\l}aczkowski},
  {Konorski}, {Villanova}, {Nardetto}, {Kervella}, {Bresolin}, {Kudritzki},
  {Storm}, {Smolec}, \& {Narloch}}]{Pietrzynski2019}
{Pietrzy{\'n}ski}, G., {Graczyk}, D., {Gallenne}, A., {et~al.} 2019, Nat, 567,
  200

\bibitem[{{Piffl} {et~al.}(2011){Piffl}, {Williams}, \&
  {Steinmetz}}]{Piffl2011}
{Piffl}, T., {Williams}, M., \& {Steinmetz}, M. 2011, A\&A, 535, A70

\bibitem[{{Placco} {et~al.}(2014){Placco}, {Frebel}, {Beers}, \&
  {Stancliffe}}]{Placco2014}
{Placco}, V.~M., {Frebel}, A., {Beers}, T.~C., \& {Stancliffe}, R.~J. 2014,
  \apj, 797, 21

\bibitem[{{Price-Whelan}(2017)}]{Price-Whelan2017}
{Price-Whelan}, A.~M. 2017, JOSS, 2, 388

\bibitem[{{Prudil} {et~al.}(2024){Prudil}, {Smolec}, {Kunder}, {Koch-Hansen},
  \& {D{\'e}k{\'a}ny}}]{Prudil2024}
{Prudil}, Z., {Smolec}, R., {Kunder}, A., {Koch-Hansen}, A.~J., \&
  {D{\'e}k{\'a}ny}, I. 2024, A\&A, 685, A153

\bibitem[{{Quispe-Huaynasi} {et~al.}(2022){Quispe-Huaynasi}, {Roig},
  {McDonald}, {Loaiza-Tacuri}, {Majewski}, {Wanderley}, {Cunha}, {Pereira},
  {Hasselquist}, \& {Daflon}}]{Quispe2022}
{Quispe-Huaynasi}, F., {Roig}, F., {McDonald}, D.~J., {et~al.} 2022, \aj, 164,
  187

\bibitem[{Ruiz-Lapuente {et~al.}(2023)Ruiz-Lapuente, Herńandez, Cartier,
  Boutsia, Figueras, Canal, \& Galbany}]{Ruiz-Lapuente2023}
Ruiz-Lapuente, P., Herńandez, J.~G., Cartier, R., {et~al.} 2023, ApJ, 947, 90

\bibitem[{{Rybizki} {et~al.}(2022){Rybizki}, {Green}, {Rix}, {El-Badry},
  {Demleitner}, {Zari}, {Udalski}, {Smart}, \& {Gould}}]{Rybizki2022}
{Rybizki}, J., {Green}, G.~M., {Rix}, H.-W., {et~al.} 2022, MNRAS, 510, 2597

\bibitem[{{Scholz}(2024)}]{Scholz2024}
{Scholz}, R.~D. 2024, A\&A, 685, A162

\bibitem[{{Sch{\"o}nrich} {et~al.}(2010){Sch{\"o}nrich}, {Binney}, \&
  {Dehnen}}]{Schonrich2010}
{Sch{\"o}nrich}, R., {Binney}, J., \& {Dehnen}, W. 2010, MNRAS, 403, 1829

\bibitem[{Silva \& Napiwotzki(2011)}]{Silva2011}
Silva, M. \& Napiwotzki, R. 2011, MNRAS, 411, 2596

\bibitem[{{Starkenburg} {et~al.}(2014){Starkenburg}, {Shetrone}, {McConnachie},
  \& {Venn}}]{Starkenburg2014}
{Starkenburg}, E., {Shetrone}, M.~D., {McConnachie}, A.~W., \& {Venn}, K.~A.
  2014, \mnras, 441, 1217

\bibitem[{Tillich {et~al.}(2009)Tillich, Przybilla, Scholz, \&
  Heber}]{Tillich2009}
Tillich, A., Przybilla, N., Scholz, R.~D., \& Heber, U. 2009, A\&A, 507, L37

\bibitem[{{Vasiliev}(2019)}]{Vasiliev2019}
{Vasiliev}, E. 2019, MNRAS, 482, 1525

\bibitem[{{Vasiliev}(2023)}]{Vasiliev2023}
{Vasiliev}, E. 2023, Galaxies, 11, 59

\bibitem[{{Vasiliev} \& {Baumgardt}(2021)}]{Vasiliev2021}
{Vasiliev}, E. \& {Baumgardt}, H. 2021, MNRAS, 505, 5978

\bibitem[{{Verberne} {et~al.}(2024){Verberne}, {Rossi}, {Koposov}, {Marchetti},
  {Kuijken}, {Penoyre}, {Evans}, {Souropanis}, \& {Tohill}}]{Verberne2024}
{Verberne}, S., {Rossi}, E.~M., {Koposov}, S.~E., {et~al.} 2024, MNRAS, 533,
  2747

\bibitem[{{Wehrhahn} {et~al.}(2023){Wehrhahn}, {Piskunov}, \&
  {Ryabchikova}}]{pysme}
{Wehrhahn}, A., {Piskunov}, N., \& {Ryabchikova}, T. 2023, \aap, 671, A171

\bibitem[{Yu \& Tremaine(2003)}]{Yu2003}
Yu, Q. \& Tremaine, S. 2003, ApJ, 599, 1129

\end{thebibliography}

\begin{appendix} 

\section{Covariance matrix} \label{Appendix}

The covariance matrix describing the uncertainties and correlations between the measured astrometric parameters ($\rho$) is:\\

$\Sigma =
\begin{pmatrix}
\sigma_{\mu_{\alpha}}^2 & \sigma_{\mu_{\alpha}} \sigma_{\mu_\delta} \rho (\mu_{\alpha}, \mu_\delta) & \sigma_{\mu_{\alpha}} \sigma_\varpi \rho (\mu_{\alpha^*}, \varpi) \\
\sigma_{\mu_{\alpha}} \sigma_{\mu_\delta} \rho (\mu_{\alpha}, \mu_\delta) & \sigma_{\mu_\delta}^2 & \sigma_{\mu_\delta} \sigma_\varpi \rho (\mu_\delta, \varpi) \\
\sigma_{\mu_{\alpha}} \sigma_\varpi \rho (\mu_{\alpha^*}, \varpi) & \sigma_{\mu_\delta} \sigma_\varpi \rho (\mu_\delta, \varpi) & \sigma_\varpi^2
\end{pmatrix}$\\
\\
\\
where $\sigma_{\mu_{\alpha}}$, $\sigma_{\mu_\delta}$, and $\sigma_\varpi$ are the uncertainties associated with $\mu_\alpha$, $\mu_\delta$, and $\varpi$, respectively. 

\end{appendix}

\end{document}